\documentclass[apj]{emulateapj}
\usepackage{apjfonts}
\usepackage{amsbsy}

\usepackage{color}

\def\ml{M/L}

\def\wp{$w_p(r_p)$}
\def\hmpc{$h^{-1}$Mpc}

\def\mstar{$M_\ast$}

\def\hmsol{$h^{-1}$M$_\odot$}

\def\navg{\langle N\rangle_M}

\def\nsat{\langle N_{\rm sat}\rangle_M}
\def\ncen{\langle N_{\rm cen}\rangle_M}
\def\mmin{M_{\rm min}}
\def\mcut{M_{\rm cut}}
\def\msat{M_{\rm sat}}

\def\asat{\alpha_{\rm sat}}

\def\om{\Omega_m}

\def\omb{\Omega_b}
\def\s8{\sigma_8}

\def\lcdm{$\Lambda$CDM}

\def\mlmean{\langle M/L\rangle}

\def\rhocrit{\rho_{\rm crit}}

\def\x2{$\chi^2$}
\def\hmsol{$h^{-1}\,$M$_\odot$}

\def\ngavg{\bar{n}_g}

\def\NNm1{\langle N(N-1) \rangle}

\def\fsat{f_{\rm sat}}
\def\slogm{\sigma_{{\rm log}M}}
\def\m_star{M_\ast}

\def\lcdm{$\Lambda$CDM}
\def\slogm{\sigma_{{\rm log}M}}
\def\om{\Omega_m}

\def\omb{\Omega_b}
\def\s8{\sigma_8}

\def\hmpc{$h^{-1}\,$Mpc}

\def\x2{$\chi^2$}
\def\hmsol{$h^{-1}\,$M$_\odot$}

\def\wp{$w_p(r_p)$}

\def\mstar{M_\ast}
\def\mmin{M_{\rm min}}

\def\mcut{M_{\rm cut}}
\def\sigmaM{\sigma_{\log M}}
\def\navg{\langle N\rangle_M}

\def\nsat{\langle N_{\mbox{\scriptsize sat}}\rangle_M}

\def\ncen{\langle N_{\mbox{\scriptsize cen}}\rangle_M}

\def\ngavg{\bar{n}_g}

\def\NNm1{\langle N(N-1) \rangle}

\def\fsat{f_{\rm sat}}

\def\sigmaM{\sigma_{\log M}}

\def\fsat{f_{\rm sat}}

\def\mstar{M_\ast}

\def\p0{P_0(r)}

\def\msat{M_{\rm sat}}

\def\cgal{c_{\rm gal}}
\def\N{N_{200}}

\def\RN{R_{\rm N200}}
\def\Sgal{\Sigma_{\rm gal}(R_p)}
\def\fcon{f_{\rm con}}

\def\m12{M_{12}}

\def\m12{M_{12}}

\def\rs0{\hat{R}_{\rm sh}^0}
\def\mg2{Mg\,II}

\def\lstar{L_\ast}

\def\fsat{f_{\rm sat}}

\def\wp{w_p(r_p)}

\def\Sp{\Sigma_{\rm gal}(R_p)}
\def\D{\Delta}
\def\DS{\Delta\Sigma(R_p)}
\def\mrz{^{0.1}M_r}

\def\fsys{f_{\rm sys}}
\def\en{\epsilon_n}
\def\eb{\epsilon_b}
\def\er{\epsilon_r}
\def\nsatbcg{N_{\rm sat}}

\begin{document}

\title{Cosmological Constraints from Galaxy Clustering\\ and the
  Mass-to-Number Ratio of Galaxy Clusters}

\author{ Jeremy L. Tinker$^1$ Erin S. Sheldon$^2$, Risa H. Wechsler$^3$,
  Matthew R. Becker$^{4,5}$, Eduardo Rozo$^{4,5,9}$, \\Ying Zu$^6$, David H. Weinberg$^6$, Idit
  Zehavi$^7$, Michael R. Blanton$^1$, Michael T. Busha$^8$, Benjamin P. Koester$^4$}
\affil{$^1$Center for Cosmology and Particle Physics, Department of Physics, New York University, New York, NY 10013\\
  $^2$Brookhaven National Laboratory, Upton, NY 11973\\
$^3$ Kavli Institute for Particle Astrophysics \& Cosmology, Physics Department, and Stanford Linear Accelerator Center, Stanford University, Stanford, CA 94305\\
$^4$ Kavli Institute for Cosmological Physics, Chicago, IL 60637\\
$^5$ Department of Astronomy \& Astrophysics, University of Chicago, Chicago IL 6037\\
$^6$ Department of Astronomy, Ohio State University, Columbus, OH 43210\\
$^7$ Department of Astronomy \& CERCA, Case Western Reserve University, Cleveland, OH 44106\\
$^8$ Institute for Theoretical Physics, Department of Physics, University of Zurich, CH-8057, Switzerland\\
$^9$ Einstein Fellow}

\begin{abstract}

  We place constraints on the average density ($\om$) and clustering
  amplitude ($\s8$) of matter using a combination of two measurements
  from the Sloan Digital Sky Survey: the galaxy two-point correlation
  function, $\wp$, and the mass-to-galaxy-number ratio within galaxy
  clusters, $M/N$, analogous to cluster $M/L$ ratios. Our $\wp$
  measurements are obtained from DR7 while the sample of clusters is
  the maxBCG sample, with cluster masses derived from weak
  gravitational lensing. We construct non-linear galaxy bias models
  using the Halo Occupation Distribution (HOD) to fit both $\wp$ and
  $M/N$ for different cosmological parameters. HOD models that match
  the same two-point clustering predict different numbers of galaxies
  in massive halos when $\om$ or $\s8$ is varied, thereby breaking the
  degeneracy between cosmology and bias. We demonstrate that this
  technique yields constraints that are consistent and competitive
  with current results from cluster abundance studies, even though
  this technique does not use abundance information. Using $\wp$ and
  $M/N$ alone, we find $\om^{0.5}\s8=0.465\pm 0.026$, with individual
  constraints of $\om=0.29\pm 0.03$ and $\s8=0.85\pm 0.06$. Combined
  with current CMB data, these constraints are $\om=0.290\pm 0.016$
  and $\s8=0.826\pm 0.020$. All errors are $1\sigma$. The systematic
  uncertainties that the $M/N$ technique are most sensitive to are the
  amplitude of the bias function of dark matter halos and the
  possibility of redshift evolution between the SDSS Main sample and
  the maxBCG cluster sample. Our derived constraints are insensitive
  to the current level of uncertainties in the halo mass function and
  in the mass-richness relation of clusters and its scatter, making
  the $M/N$ technique complementary to cluster abundances as a method
  for constraining cosmology with future galaxy surveys.

\end{abstract}


\section{Introduction}

Galaxy clustering measurements offer a unique window into the
distribution of dark matter in the universe. The recently completed
Sloan Digital Sky Survey (SDSS; \citealt{york_etal:00}, \citealt{dr7})
provides unprecedented precision and accuracy in its map of the local
universe.  However, one of the main impediments to the use of galaxy
clustering for inferring cosmological information is galaxy bias---it
is probable that the distribution of galaxies differs from the
distribution of dark matter. Thus both the amplitude and shape of the
galaxy clustering signal are biased relative to the clustering of dark
matter at quasi-linear and nonlinear scales. This bias is degenerate
with cosmology in such a way that a bias model can be constructed to
match the observed real-space galaxy two-point correlation function
for a range of cosmological models (\citealt{yang_etal:04_mocks,
  tinker_etal:05, yoo_etal:06, zheng_weinberg:07}). In this paper, we
combine real-space galaxy clustering and the cross-correlation of
galaxies with galaxy clusters. The mass scale of the clusters is
determined by weak gravitational lensing from
\cite{sheldon_etal:09_data}, which provides the leverage to break the
degeneracy and constrain cosmological parameters.

The framework within which we construct the bias model is the Halo
Occupation Distribution (HOD; see, e.g., \citealt{peacock_smith:00,
  seljak:00, roman_etal:01, berlind_weinberg:02, cooray_sheth:02}). In
this model, all galaxies are contained within dark matter halos with a
probability distribution $P(N|M)$, the probability that a halo of mass
$M$ constrains $N$ galaxies of a given class. We define a dark matter
halo as a collapsed, virialized object with a mean interior density of
200 times the critical density. The statistics governing the spatial
distribution of dark matter can be estimated with analytic models
(e.g., \citealt{press_schechter:74, cole_kaiser:89, mo_white:96,
  sheth_tormen:99, smt:01, ma_etal:10}), but they are most accurately
calibrated through $N$-body simulations (\citealt{jenkins_etal:01,
  seljak_warren:04, warren_etal:06, reed_etal:07, tinker_etal:08_mf,
  pillepich_etal:10, tinker_etal:10_bias}). These simulations specify
the mass function of dark matter halos, the large-scale bias of halos,
and their quasi-linear clustering relative to the clustering of the
matter distribution itself. Thus, once the HOD is constructed with a
model for the internal distribution of galaxies within halos, this
relationship between galaxies and halos fully specifies the spatial
distribution of galaxies on all scales. The key ingredient within
$P(N|M)$ is the mean number of galaxies as a function of halo mass,
$\navg$. Distinct cosmologies produce distinct populations of dark
matter halos (\citealt{zheng_etal:02}), thus for each cosmology there
is a distinct $P(N|M)$ and $\navg$ such that the model matches the same
galaxy two-point correlation function $\xi_g(r)$.

Data sets that are sensitive to the underlying mass scale of dark
matter halos can break the bias-cosmology degeneracy
(\citealt{zheng_weinberg:07}). The most direct measurement is the mean
number of galaxies in halos of mass $M$. In this paper we will use the
ratio of these two quantities, $M/N$. This measurement is most easily
made in the most massive halos---galaxy clusters. Cluster-sized halos
are the largest collapsed structures in the dark matter density
distribution, making them relatively easy to locate observationally
and providing myriad methods with which to estimate their masses. The
$M/N$ measurement is analogous to the mass-to-light ratio of dark
matter halos, $M/L$. In previous decades, $M/L$ of galaxy clusters was
utilized as a method for inferring the dark matter density parameter
$\om$ (e.g., \citealt{gott_etal:74, peebles:86, bahcall_etal:95,
  bahcall_etal:00, carlberg_etal:96, rines_etal:04}). The pivotal
assumption in this method was that the mean $M/L$ of clusters was
representative of the mean $M/L$ of the universe, thus $\om=\mlmean
\times \rho_{\rm lum}/\rhocrit$, where $\rho_{\rm lum}$ is the
luminosity density and $\rhocrit$ is the critical density. These
results were among the first to challenge the prevailing theoretical
expectation that $\om=1$.  However, efforts to use this relation
yielded values of $\om$ near 0.1, in significant tension with a large
array of other methods that were converging on a `concordance'
cosmology of $\om=0.3$ and $\s8=0.9$ (e.g., \citealt{spergel_etal:03,
  tegmark_etal:04, seljak_etal:05}). However appealing, the assumption
that clusters are representative had little theoretical support
(\citealt{bbks}), and the lack of agreement between cluster $\ml$ and
other measures of $\om$ strongly implied that they are biased objects.

In \cite{tinker_etal:05}, we demonstrated that cluster $M/L$ ratios depend
not only on the matter density but also on the amplitude of density
fluctuations, $\s8$, once the galaxy bias model is constrained to
match the projected galaxy correlation function
$\wp$. Combining $M/L$ measurements of clusters in the CNOC2 survey
(\citealt{carlberg_etal:96}) with early clustering results from SDSS
(\citealt{zehavi_etal:05}), we determined that these data constrain a
degeneracy curve $(\om/0.3)^{0.6}(\s8/0.9)=0.75\pm 0.06$. This
degeneracy is similar to that constrained by measurements of the
abundances of galaxy clusters; `cluster normalized' models that lie on
the same value of $\om^{\gamma}\s8$ predict the same number of
clusters, where $\gamma$ is usually around $0.4-0.6$
(\citealt{rozo_etal:10}). These results were in good agreement with
those of \cite{vdb_etal:03}, who combined clustering in the Two Degree
Field Galaxy Redshift Survey (\citealt{norberg_etal:01}) with cluster
$M/L$ and first-year CMB results from WMAP
(\citealt{spergel_etal:03}). \cite{vale_ostriker:06} reached similar
conclusions using a subhalo abundance matching method, again drawing
on cluster $M/L$ ratios. All of these studies demonstrated that the
`concordance' values of $(\om,\s8)=(0.3,0.9)$ could not simultaneously
account for the clustering of galaxies and the $M/L$ of
clusters. These conclusions were validated when three-year WMAP
results emerged in excellent agreement in the $\om$-$\s8$ plane
(\citealt{spergel_etal:07}) with the $M/L$ result of
$(\om/0.3)\times(\s8/0.9)^{0.6}=0.75\pm 0.06$ from
\cite{tinker_etal:05}. The key revision to the WMAP estimates was
improved correction of polarization foregrounds, which reduced the
estimated electron scattering optical depth, and which thereby reduced
the matter fluctuation amplitude inferred from the CMB anisotropy
amplitude.

Although the confirmation with later CMB data demonstrated the
potential of this approach, both the \cite{vdb_etal:03} and
\cite{tinker_etal:05} studies paid little attention to the possible
systematic errors in the use of halo occupation methods, emphasizing
instead the systematic uncertainties in cluster $M/L$ estimates.
this paper, we use the largest data sets currently available. For the
clustering we use measurements from Data Release 7 of the SDSS
(\citealt{dr7}), which covers nearly a quarter of the sky to obtain
close to a million galaxy redshifts. Our cluster sample comes from the
maxBCG cluster catalog (\citealt{koester_etal:07}), consisting of over
$10^4$ optically-detected clusters. These massive data sets close the
door on statistics-limited analyses; in this paper we take great care
to quantify the systematic uncertainties in our theoretical model. Our
knowledge of the statistics of dark matter halos is not perfect, and
our assumptions for how galaxies populate halos are not
infallible. Thus we will incorporate these uncertainties into our
analysis in order to yield robust constraints on cosmological
parameters.

This paper is structured as follows: In section 2 we present our
clustering data and our measurements of $M/N$ from the maxBCG
clusters. In section 3 we describe our theoretical model for both
$\wp$ and $M/N$ within the HOD framework, demonstrating how this
method is sensitive to different cosmologies. In section 4 we present
a detailed description of the systematic uncertainties in our
analysis. In section 5 we present our results, showing constraints
in the $\om$-$\s8$ plane and constraints on HOD parameters. In
section 6 we discuss these results in context with other studies that
utilize clusters as a cosmological tool and present prospects for the
future with the $M/N$ method. We briefly summarize in section 7.

\begin{figure*}
\epsscale{1.0} 
\plotone{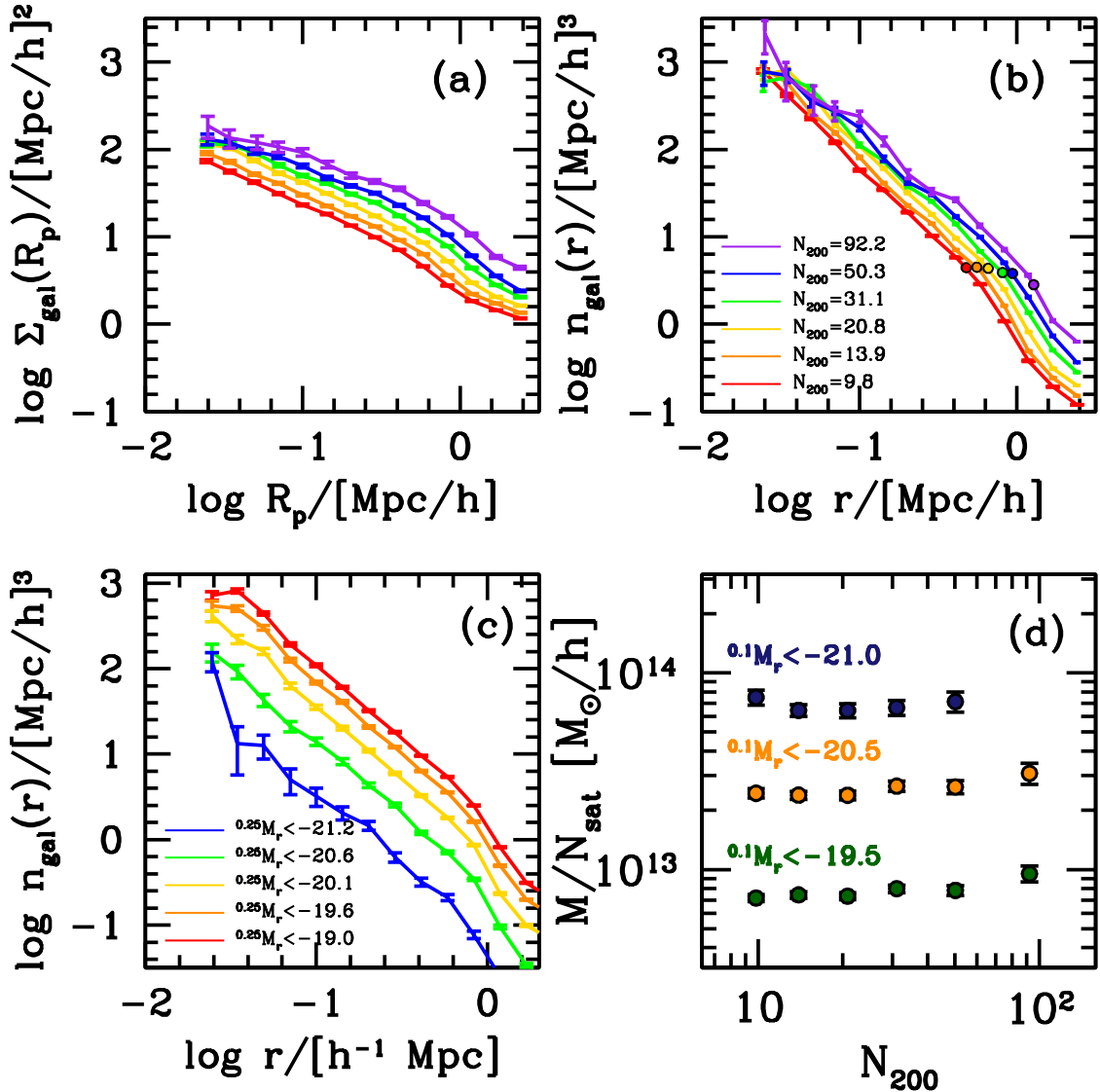}
%\vspace{-7.5cm}
\caption{ \label{maxbcg_data} Panel (a): The background-subtracted
  projected galaxy number density around maxBCG clusters of varying
  richness (see key in panel [b]). These measurements are for galaxies
  brighter than $^{0.25}M_r=-19.0$. Panel (b): The deprojected number
  density profiles of galaxies for the data in panel (a). The filled
  circle indicates the halo radius $\RN$ for each richness bin. Panel
  (c): The variation in $n_{\rm gal}(r)$ as a function of magnitude
  threshold. Panel (d): The $M/N$ measurements as a function of
  cluster richness. The data are labeled by their $\mrz$ magnitude
  thresholds (see Table 1 for the conversion between $\mrz$ and
  $^{0.25}M_r$). Note that we define $N$ to include on satellite
  galaxies. These data do not include the correction for miscentering,
  discussed in \S 4. }
\end{figure*}

\section{Data}

\subsection{Galaxy Clustering Measurements}
\label{s.wp_data}

Our galaxy clustering measurements are obtained from 7300 deg$^2$ of
sky, slightly smaller than the full SDSS DR7 of 7900 deg$^2$
(\citealt{dr7}). The methodology is described in detail by
\citeauthor{zehavi_etal:10} (2010; hereafter Z10). Briefly, we choose
three volume-limited samples of galaxies on which to focus our
analysis: $\mrz<-19.5$, $\mrz<-20.5$, and $\mrz<-21.0$ (also listed in
Table 1)\footnote{We do not assume a value of $h$ for galaxy
  magnitudes, but for brevity we write absolute magnitudes as $M$
  rather than $M-5\log h$. The superscript indicates the redshift to
  which the magnitude is $k$-corrected using the algorithm of
  \cite{blanton_roweis:07}.}. These samples have complementary
attributes for this analysis. The $M_r<-20.5$ sample, which
corresponds approximately to a sample of all galaxies brighter than
$\lstar$, is the optimal sample for this type of analysis; the
magnitude limit is bright enough such that the volume contained
suppresses sample variance in the clustering measurement, but it has a
high enough number density that the $\navg$ will be above unity for
the mass range probed in the sample of clusters we use. The fainter
sample, $M_r<-19.5$ maximizes the signal-to-noise in the $M/N$
measurement, while the brighter sample, $M_r<-21.0$, has the smallest
sample variance for the clustering measurement. The galaxy clustering
measurements are only weakly correlated because $z_{\rm max}$ is
different for each sample, minimizing the overlapping volume (see
Table 1).

The clustering quantity we utilize for each sample is the projected
two-point galaxy autocorrelation function, $\wp$. This statistic is
defined as

\begin{equation}
\label{e.wp}
\wp = 2\int_0^{\pi_{\rm max}}\xi(r_p, \pi) d\pi,
\end{equation}

\noindent where $r_p$ is the projected separation between two
galaxies, $\pi$ is the line-of-sight separation between two galaxies,
and $\xi(r_p,\pi)$ is the measured two-dimensional correlation
function. Due to the peculiar motions of galaxies against the Hubble
flow, $\xi(r_p,\pi)$ is anisotropic. Integrating along the $\pi$
direction minimizes the effects of redshift space anisotropies,
allowing for an easier comparison to analytic models. If $\pi_{\rm
  max}=\infty$, redshift space effects are eliminated entirely, but
the finite volume of the survey requires that we limit the integral to
$\pi_{\rm max}=40$ \hmpc. At this $\pi_{\rm max}$, redshift-space
effects on $\wp$ are larger than the error bars on the DR7
measurements at scales of $r_p\gtrsim 5$ \hmpc. Although it is
possible to include redshift-space anisotropies in HOD calculations of
the non-linear two-point correlation function (e.g.,
\citealt{tinker:07}), we limit our analysis to $r_p<3$ \hmpc\ to
eliminate any systematic biases due to redshift-space effects. Our
theoretical model for $\wp$ utilizes the \cite{smith_etal:03} fitting
function for the nonlinear matter $\xi(r)$, thus limiting our analysis
to small scales eliminates biases from possible errors in the
\cite{smith_etal:03} prescription.  Measurements at these scales
provide more than enough statistical power for our constraints given
the systematic uncertainties discussed in \S 4, which dominate our
final error budget.

We use the full covariance matrix for each clustering sample. The
covariance matrix is estimated through the data using the jackknife
technique. The full survey is divided into 104 subsamples of roughly
equal sky area (thus nearly equal volume), and each subsample is
removed from the survey and $\wp$ is re-measured. The jackknife
cumulative covariance estimate is

\begin{equation}
\label{e.covar}
\sigma^2_{ij} = \frac{N-1}{N}\sum_{l=1}^N\left(w_{p,i}^l-\bar{w}_p\right)\left(w_{p,j}^l-\bar{w}_p\right),
\end{equation}

\noindent where subscripts $i$ and $j$ denote bins of $r_p$ and
superscript $l$ denotes the subsample. The factor of $(N-1)/N$
corrects for the fact that the variance is calculated with samples
that are slightly smaller than the full sample, thus the variance of
the subsamples will be larger approximately by the ratio of the
volumes. The accuracy of these error estimates is discussed by Z10;
see also \cite{norberg_etal:09}, who conclude that jackknife errors
are generally accurate or conservative on the scales 
investigated here.

\begin{figure}
\epsscale{1.0} 
\plotone{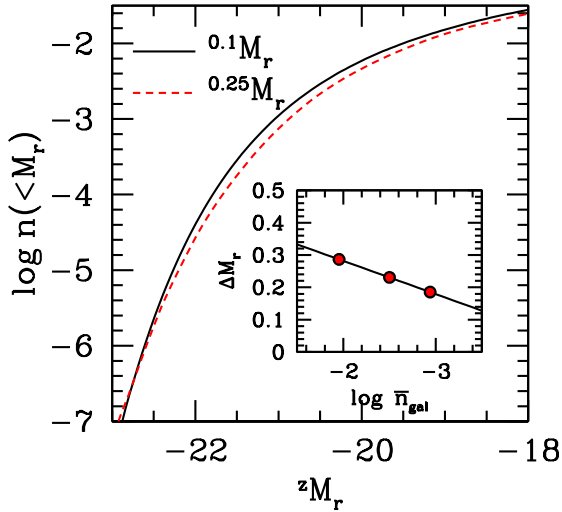}
%\vspace{-7.5cm}
\caption{ \label{lumfunc_evolution} The evolution of the $r$-band
  cumulative luminosity function between $z=0.1$ and $z=0.25$. The
  black curve shows the measurements from \cite{blanton_etal:03}. The
  dashed curve represents our estimate of the $^{0.25}M_r$ luminosity
  function based on the method of \cite{blanton:06}. The inset panel
  shows $\Delta M_r = ^{0.25}M_r - ^{0.1}M_r$. The three filled points
  indicate the number densities of our three $z=0.1$ Main galaxy
  samples. See Table 1 for the exact values of $^{0.25}M_r$ for the
  three $\mrz$ thresholds.}
\end{figure}

\subsection{The maxBCG Cluster Catalog}

The maxBCG algorithm (\citealt{koester_etal:07_methods}) utilizes the
high fraction of early-type galaxies within clusters to identify
massive halos as overdensities of bright, uniformly red galaxies. The
large number of such galaxies within each cluster and the tightness of
their color distribution allow for accurate photometric redshift
estimates. Using mock galaxy catalogs, the algorithm is optimized to
have completeness and purity above 90\%, or higher as richness
increases. The algorithm produces a richness estimate for each object,
$\N$, that is defined as the number of red-sequence galaxies brighter
than $^{0.25}M_i=-19.2$ within an aperture containing a mean
overdensity of such galaxies that is 200 times the mean density of
galaxies. This aperture scales roughly with the virial radius of the
halo, thus larger halos have larger apertures
(\citealt{hansen_etal:05}). The richness estimate is not to be
confused with the occupation number in the HOD; $N$ in the HOD refers
to {\it all} the galaxies (not just red) brighter than a defined
magnitude threshold within the exact halo radius. The richness $\N$,
although correlated with $N$ to some degree, is used exclusively as a
reference value with which to bin clusters. Because the magnitude
limit for maxBCG selection is significantly fainter than that of even
our $^{0.1}M_r < -19.5$ sample, and the aperture within which maxBCG
counts member galaxies is not identical to the actual halo radius, the
values of $N$ are lower than those of $N_{200}$, by roughly a factor
of two for $^{0.1}M_r < -19.5$ and larger factors for brighter
thresholds (see Table 2 below; also recall that $N_{200}$ includes
central galaxies while $N$ counts only satellites).

The maxBCG catalog (\citealt{koester_etal:07}) is a sample of clusters
identified in 7398 deg$^2$ of SDSS imaging data, roughly the same as
the imaging samples released in DR4 (\citealt{dr4}). For our analysis
here, we utilize a subset of this imaging data that was analyzed in
\cite{sheldon_etal:09_data} and \cite{sheldon_etal:09_ml} consisting
of 6325 deg$^2$. The redshift range of the clusters in the sample is
$0.1\le z\le 0.3$, yielding a nearly volume-limited sample of
clusters. Although the catalog identifies clusters with $\N\ge3$, we
restrict our analysis to clusters with $\N\ge 9$. Objects with a
richness below this value suffer from a higher degree of uncertainty
due to projection effects. We bin the clusters into 6 richness bins
listed in Table 2. We note that our catalog differs slightly from the
publicly released catalog in \cite{koester_etal:07}, which was
limited to clusters with $\N\ge 10$.

The average masses of clusters in each richness bin are obtained from
the weak gravitational lensing analysis of
\cite{sheldon_etal:09_data}. All the clusters in each bin are stacked,
yielding a high signal-to-noise estimate of the projected density
contrast profile, $\DS$. This profile is deprojected and integrated
out to a radius at which the mean interior density is
$\D=200\rhocrit$, thereby yielding an estimate of the mean cluster
mass within each bin. We define this radius as $\RN$. We differentiate
this radius from $R_{200c}$, which is the radius for a specific halo,
while $\RN$ is a quantity specific to each richness bin\footnote{Note
  that $\RN$ is different from the quantity $R_{200}$, defined in
  \cite{hansen_etal:05} as the radius at which the mean interior
  {\it galaxy} density within clusters is 200 times the mean galaxy
  density.}. The weak lensing measurements and additional tests with
$X$-ray measurements and velocity dispersions of maxBCG clusters
demonstrate that $\N$ correlates strongly with dark matter mass, but
there is also a scatter between mass and richness
(\citealt{becker_etal:07, rykoff_etal:08, rozo_etal:09_scatter}). This
scatter and its uncertainties are characterized by
\cite{rozo_etal:09_scatter}. 

The cluster masses in \cite{sheldon_etal:09_data} are subject to a
number of biases. First, this analysis assumes that brightest cluster
galaxy (BCG) identified by the algorithm is located at the true bottom
of the gravitational potential of the dark matter halo. Tests on mock
galaxy catalogs reveal that this is not true in $\sim 10\%$ of
observed objects, a fraction that increases monotonically with
decreasing richness (\citealt{johnston_etal:07}). This `miscentering
effect' lowers the measured mass relative to the true
mass. Additionally, errors in the photometric redshifts of the sample
of background sources behind the clusters also affect the measured
halo mass (see the discussion in \citealt{rozo_etal:09_scatter}). The
uncertainties in scatter, miscentering, and weak lensing systematics
are all taken into account in this study. In particular, the
\cite{sheldon_etal:09_data} weak lensing masses are estimated to be
biased low by $18\pm 6\%$ (\citealt{rozo_etal:09_scatter}).  We will
discuss their quantitative incorporation in the analysis in \S
\ref{s.systematics}. One advantage of our $M/N$ analysis is that it is
much less sensitive than abundance analysis to miscentering errors,
since these tend to affect mass and galaxy occupation in similar ways.

\subsection{Measuring the Number of Galaxies in Clusters}

To determine the number of satellite galaxies per halo, a stacking
technique is also used. In each stacked richness bin, the total
projected galaxy counts are measured in bins of $\log R_p$. To
determine the number of galaxies that are associated with the cluster
and not chance projection, the same process of stacking is done with a
set of random pointings with the same redshift distribution as the
clusters. In each bin of $\log R_p$, the mean number of galaxies from
the random projections is subtracted from the mean number around the
stacked clusters. Full details of this procedure are given by
\cite{sheldon_etal:09_ml} and \cite{hansen_etal:09}.

Figure \ref{maxbcg_data}a shows the projected galaxy number density
profiles for the six richness bins in Table 2. These results are shown
for galaxies brighter than $^{0.25}M_r=-19.0$. The density profiles do
not include the BCG of each cluster, which is located at $R_p=0$. Thus
these data represent satellite galaxies only. The error bars are
obtained through jackknife resampling in the plane of the sky with
patches of $6.3$ deg$^2$. To obtain the number of galaxies within
$\RN$, $\Sgal$ is inverted to recover the three-dimensional density
profile. The inversion is performed using the standard Abel-type
integral,

\begin{equation}
\label{e.abel}
n_{\rm gal}(r) = \frac{1}{\pi} \int_r^\infty dR_p \frac{-\Sp}{\sqrt{r^2-R_p^2}}.
\end{equation}

\noindent The projected density is not measured out to infinity, thus
in practice we fit a power law to the three highest-$R_p$ data points
and truncate the integral at 30 \hmpc. Making this upper limit twice
as large results in negligible differences in the density profiles
obtained within the cluster radii. Figure \ref{maxbcg_data}b
shows the inverted three-dimensional density profiles of the satellite
galaxies. The filled circles indicate the cluster radius in each
richness bin. These density profiles are integrated out to $\RN$ to
determine the total halo occupation of satellite galaxies in each
richness bin. Figure \ref{maxbcg_data}c shows how the
number density profiles depend on luminosity threshold for a fixed
richness bin, $\langle N_{200}\rangle=20.8$ in this example. As
expected, the number of galaxies increases monotonically with
decreasing luminosity threshold at all scales.

The clustering measurements described in \S \ref{s.wp_data} are for
the Main sample of SDSS galaxies, which probe $z\sim 0.1$. The median
redshift of the maxBCG catalog is $z\sim 0.25$. For proper analysis,
we require a consistent sample of galaxies between both
redshifts. Because the luminosity function evolves between these two
redshifts, and because $^{0.25}M_r$ is not equivalent to $\mrz$, we
choose to define $z=0.25$ samples that have the same number density as
the Main samples. Figure \ref{lumfunc_evolution} compares the
$^{0.1}M_r$ luminosity function (\citealt{blanton_etal:03}) to our
estimate of the $^{0.25}M_r$ luminosity function using the technique
of \cite{blanton:06} and employed in the \cite{sheldon_etal:09_ml}
analysis. The \cite{blanton:06} technique uses measurements of the
luminosity function at $z\sim 0.1$ and $z\sim 1$ to constrain the
amount of luminosity evolution and number density evolution in both
the red sequence and the blue cloud across this redshift
baseline. These results are used constrain simple star formation
histories of blue and red galaxies that are used to interpolate the
total galaxy luminosity function at $z=0.25$. The inset panel shows
the magnitude shifts between $^{0.1}M_r$ and $^{0.25}M_r$ at the three
$z=0.1$ magnitude thresholds. We will discuss the systematic
uncertainties of this approach in \S \ref{s.systematics}.

\begin{figure}
\epsscale{2.5} 
\plotone{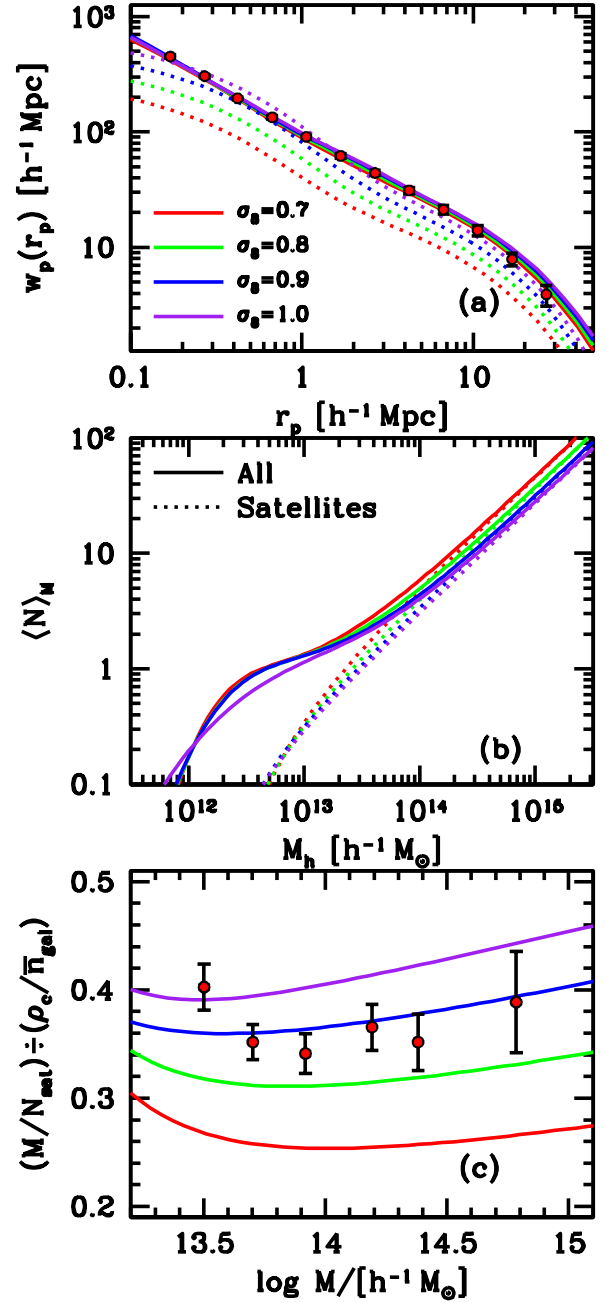}
%vspace{-12cm}
\caption{ \label{hod_demo} A pedagogical demonstration of the $M/N$
  method. Panel (a): The circles show the projected correlation
  function, $\wp$, measured for $\mrz<-20.5$ galaxies in DR7. The
  solid curves show HOD model fits to these data; in each model, the
  value of $\s8$ is changed. The dotted lines correspond to the
  projected correlation function for the dark matter in each
  cosmological model. Panel (b): The occupation functions inferred
  from the best-fit models to $\wp$ for each value of $\s8$. As $\s8$
  decreases, the amplitude of $\navg$ at high masses increases. Panel
  (c): The points with errors show the maxBCG measurements for $M/N$
  (now with the miscentering correction from Figure
  \ref{miscentering_correction} applied). The curves show the
  predictions for this quantity for each value of $\s8$. For
  convenience, we have plotted $M/N$ as a function of halo mass rather
  than cluster richness, though in actual model fitting we use
  richness. }
\end{figure}

\begin{figure}
\epsscale{1.0} 
\plotone{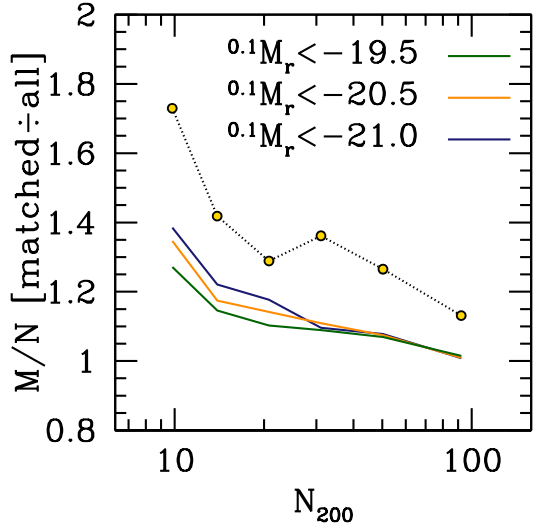}
%\vspace{-7.5cm}
\caption{ \label{miscentering_correction} The effect of cluster
  miscentering on $M/N$ in mock galaxy catalogs. The solid curves show
  the ratio of $M/N$ measured from a sample of clusters that are
  well-matched to halos in the mock catalog to $M/N$ measured on the
  full sample of clusters determined in the mock. Miscentering lowers
  both the mass and the number of galaxies within clusters, but the
  effect largely cancels in the ratio. The dotted curve connecting the
  open circles indicates the effect of miscentering on the halo masses
  derived from the $\DS$ measurements in \cite{johnston_etal:07}.  In
  the mock, both $M$ and $N$ are measured from the projected mass and
  number density, respectively, around the clusters detected in the
  mock. We use the mean of all three curves as the miscentering
  correction on the data, and use the maximal difference of the three
  curves as the error in that correction.}
\end{figure}

The error bars on $n_{\rm gal}(r)$ are correlated, thus to be
conservative we assume that the errors are 100\% correlated; we
determine the errors on $\nsatbcg(<M^{\rm lim}_r)$ by increasing or
decreasing every $n_{\rm gal}(r)$ datum by its $1\sigma$ error and
re-integrating. Our $n_{\rm gal}(r)$ data are measured on a grid of
threshold $^{0.25}M_r$ values spaced by 0.28 magnitudes. For each
threshold, we integrate $n_{\rm gal}(r)$ out to $\RN$ to obtain
$N_{\rm sat}$ and interpolate to obtain $N_{\rm sat}$ for the exact
values of $^{0.25}M_r$ that yields the same number density as the
$\mrz$ samples (cf. Figure \ref{lumfunc_evolution}). The values and
uncertainties of $\nsatbcg(<M^{\rm lim}_r)$ for our three magnitude
threshold values are listed in Table 2. Figure \ref{maxbcg_data}d
shows the resulting $M/N$ ratio for our three $^{0.1}M_r$ magnitude
thresholds. Note that the $N$ in $M/N$ here and throughout this work
refers to the number of satellite galaxies, a choice that makes $M/N$
approximately constant from bin to bin, which would not be the case if
we included central galaxies. To determine the
statistical error on $M/N$, we sum in quadrature the fractional errors
on $\nsatbcg$ and $M$ from the weak lensing measurements in
\cite{sheldon_etal:09_data}. We have 17 total $M/N$ data points, 6 for
the $\mrz<-19.5$ and $\mrz<-20.5$ samples, and 5 for the $\mrz<-21.0$
sample. For the brightest sample, the richest bin does not have enough
clusters to obtain a robust density profile. Values of our $M$ and $N$
measurements for the maxBCG sample are listed in Table 3.

When comparing $M/N$ measurements to model predictions, we must
account for the fact that the $M$ in the $M/N$ ratio is the same value
for all three magnitude threshold samples. Thus the errors are
correlated between the samples. We define a covariance matrix for the
17 $M/N$ data points where $i=1-6$ represent the $\mrz<-19.5$ sample,
$i=7-12$ represents $\mrz<-20.5$, and $i=13-17$ represents
$\mrz<-21$. Most of the off diagonal elements of the covariance matrix
are zero, but, for example, data points 1, 7, and 13 are correlated
because the halo mass is the same for these three data points. The
diagonal elements of the covariance matrix are obtained by total
differential error on the $M$ and $N$ data,

\begin{equation}
C_{ii} = \sigma_{Mi}^2\left(\frac{\partial(M/N)_i}{\partial M_i} \right)^2
  + \sigma_{Ni}^2\left(\frac{\partial(M/N)_i}{\partial N_i}\right)^2.
\end{equation}

\noindent The off-diagonal terms are also calculated in the same
manner. For example, $C_{1,7}$ is a non-zero element expressed by

\begin{equation}
C_{1,7} = \sigma_M^2 \frac{\partial(M/N)_1}{\partial M} 
\frac{\partial(M/N)_7}{\partial M},
\end{equation}

\noindent where $M$ for the two $M/N$ data points is the same
value. We note that accounting for this covariance makes little
difference in our results; the uncertainties in the $M/N$ measurements
are dominated by systematics that we will discuss in \S 4.

As can be seen in Figure \ref{maxbcg_data}d, the $M/N$ ratio is
roughly independent of cluster richness. In most models of halo
occupation, the number of satellite galaxies scales as a power-law
with host halo mass. Most results from observed galaxy clustering, as
well as N-body simulations, yield a power-law index close to unity,
which would imply $M/N\sim$ constant (\citealt{kravtsov_etal:04,
  zheng_etal:05, conroy_etal:06, vdb_etal:07, tinker_etal:07,
  zheng_etal:07, yang_etal:08}). 

%%%%%%%%%%%%%%%%%%%%%%%%%%%%%%%%%
% Table 1
%%%%%%%%%%%%%%%%%%%%%%%%%%%%%%%%%
\begin{deluxetable*}{cccccc}
  \tablecolumns{7} \tablewidth{25pc} \tablecaption{SDSS DR7 Spectroscopic Clustering Samples}
  \tablehead{\colhead{$^{0.1}M_r$} & \colhead{$^{0.25}M_r$} & \colhead{$z_{\rm med}$} & \colhead{$z_{\rm max}$ } & \colhead{$N_{\rm gal}$} & \colhead{$\ngavg$/(\hmpc)$^{-3}$} }

\startdata

-19.5 & -19.21 & 0.068 & 0.083 & 112497 & $1.11\times10^{-2}$ \\
-20.5 & -20.27 & 0.104 & 0.131 & 117588 & $3.16\times10^{-3}$ \\
-21.0 & -20.81 & 0.126 & 0.160 & 77381 & $1.15\times10^{-3}$  \\

\enddata \tablecomments{The number density is corrected for
  incompleteness. $^{0.25}M_r$ is the magnitude threshold at $z=0.25$
  with the same number density as the corresponding $\mrz$.}
\end{deluxetable*}
%%%%%%%%%%%%%%%%%%%%%%%%%%%%%%%%%
% END Table 1
%%%%%%%%%%%%%%%%%%%%%%%%%%%%%%%%%

%%%%%%%%%%%%%%%%%%%%%%%%%%%%%%%%%
% Table 2
%%%%%%%%%%%%%%%%%%%%%%%%%%%%%%%%%
\begin{deluxetable*}{llllllll}
  \tablecolumns{8} 
  \tablewidth{40pc} 
  \tablecaption{maxBCG Cluster Sample} 
  \tablehead{\colhead{$\N$ range} & \colhead{$M_{\rm N200}/10^{13}$ \hmsol} & \colhead{$N(\mrz<-19.5)$} & \colhead{$N(\mrz<-20.5)$} & \colhead{$N(\mrz<-21)$} & \colhead{miscentering} & \colhead{$\RN$ \hmpc} }

\startdata

9-11 & $3.17\pm 0.12$ & $4.06 \pm 0.08$ & $1.25\pm 0.04$ & $0.450 \pm 0.029$ & $1.33\pm 0.06$ & 0.59 \\
12-17 & $5.04\pm 0.16$ & $6.21 \pm 0.11$ & $2.00\pm 0.06$ & $0.772\pm 0.042$ & $1.18\pm 0.04$ & 0.70 \\
18-25 & $8.28 \pm 0.30$ & $10.40 \pm 0.22$ & $3.31\pm 0.13$ & $1.258\pm 0.082$ & $1.14\pm 0.04$ & 0.82 \\
26-40 & $15.54\pm 0.59$ & $17.81 \pm 0.41$ & $5.57\pm 0.23$ & $2.26\pm 0.16$ & $1.098\pm 0.009$ & 1.01 \\
41-70 & $24.0\pm 1.1$ & $27.92 \pm 0.85$ & $8.78\pm 0.48$ & $3.40 \pm 0.31$ & $1.074\pm 0.004$ & 1.17 \\
71-220 & $60.9\pm 4.5$ & $58.0 \pm 3.1$ & $18.8\pm 1.7$ & --- &  $1.011\pm 0.003$ & 1.60 \\

\enddata \tablecomments{Note: The halo masses are equivalent to those
  in \cite{sheldon_etal:09_ml} with an 18\% correction factor as
  discussed in \cite{rozo_etal:09_scatter}. The $N$ values refer only
  to satellite galaxies and are given for the equivalent $^{0.1}M_r$
  thresholds; see text for details. The values listed are the raw
  measurements that do not include the correction for
  miscentering. The `miscentering' column indicates the factor, and
  its error, by which all $N$ measurements are multiplied to correct
  for miscentering.  }
\end{deluxetable*}
%%%%%%%%%%%%%%%%%%%%%%%%%%%%%%%%%
% END Table 2
%%%%%%%%%%%%%%%%%%%%%%%%%%%%%%%%%

\section{Theoretical Modeling}

\subsection{Galaxy Two-Point Correlation Function}
\label{s.hod_model}

We parameterize the halo occupation function as two separate
functions, one representing the occupation of central galaxies and one
for the occupation of satellite galaxies. For central galaxies, we use
the standard expression

\begin{equation}
\label{e.ncen}
\ncen = \frac{1}{2}\left[ 1+\mbox{erf}\left(\frac{\log M - \log
    \mmin}{\sigmaM} \right) \right],
\end{equation}

\noindent where $\mmin$ formally represents the mass at which a halo
has a 50\% probability of containing a central galaxy bright enough to
be contained within the sample. The parameter $\slogm$ is related to
the scatter in halo mass at fixed luminosity. Functionally, this
parameter controls how ``sharp'' the transition is between halos that
host no galaxies and halos that have one central galaxy. Because halo
mass is monotonically related to clustering strength over most of the
halo mass spectrum, the value of $\slogm$ correlates with the
large-scale bias of the model; a higher scatter brings more low-mass
halos into the sample, and due to the steepness of the halo mass
function the resulting large-scale bias is reduced. When $M\lesssim
\mstar$, bias is relatively independent of halo mass\footnote{We
  define the non-linear mass scale $\mstar$ as the mass at which the
  linear matter variance on the Lagrangian scale of the halo is
  $\sigma(M)=1.686$.}, thus $\slogm$ has little effect on $\wp$. For
this reason, we allow $\slogm$ to be a free parameter for the
$\mrz<-20.5$ and $\mrz<-21$ samples, but we fix it at $\slogm=0.2$ for
the faint $\mrz<-19.5$ sample. In our modeling we adopt a flat prior
of $0.05<\slogm<1.6$. Values of $\slogm<0.05$ are indistinguishable
from $\slogm=0$, while values of $\slogm>1.6$ would be unphysical.

For satellite galaxies, we adopt an occupation function of the form

\begin{equation}
\label{e.nsat}
\nsat = \ncen\times \left(\frac{M}{\msat}\right)^{\asat}\exp\left(\frac{-\mcut}{M}\right).
\end{equation}

\noindent Equation (\ref{e.nsat}) parameterizes satellite occupation
as a power-law at high halo masses with an exponential cutoff at low
masses, motivated by the results of high-resolution N-body and
hydrodynamic simulations of galaxy formation
(\citealt{kravtsov_etal:04, zheng_etal:05, conroy_etal:06,
  wetzel_white:10}). The factor of $\ncen$ in equation (\ref{e.nsat})
ensures $\nsat\le\ncen$ at all masses.

We assume that satellite galaxies follow a spatial distribution
within the dark matter halo of an NFW density profile
(\citealt{nfw:97}). However, we do {\it not} assume that the galaxies
trace the dark matter within the halos. The concentration parameter of
NFW profile, $\cgal$, is a multiple of the concentration parameter of
the dark matter, $\fcon\equiv\cgal/c_m$. The proportionality constant
$\fcon$ is left as a parameter in our analysis with a flat prior of
$[0.2,2.0]$, which brackets the extreme scenarios for the spatial bias
between satellite galaxies and dark matter within halos. Thus the {\it
  shape} of the concentration-mass relation is the same as the dark
matter, but the normalization is allowed to vary.  For the dark
matter, we use the concentration-mass relation of
\cite{bullock_etal:01} with the parameters of \cite{wechsler_etal:06}.

To calculate the mean number of pairs within each halo, we assume that
the satellite galaxies are Poisson distributed about the mean. This is
well supported by both numerical results (\citealt{kravtsov_etal:04,
  zheng_etal:05, gao_etal:11}) and observational data
(\citealt{lin_etal:04}). Small deviations from Poisson like those shown
in the recent results from \cite{busha_etal:10} and
\cite{boylan_kolchin_etal:10} do not have a significant effect on the
clustering in this paper because they occur at subhalo masses not
probed for the galaxy samples analyzed here.

We use the theoretical model of \cite{tinker_etal:05} to calculate the
two-point correlation function in real space, with one
modification. Because the mass function and bias relation used in this
analysis are taken from numerical results based on
spherical-overdensity (SO) halo catalogs (\citealt{tinker_etal:08_mf,
  tinker_etal:10_bias}), the halo exclusion must be modified to match
this halo definition. In the SO halo finding algorithm of
\cite{tinker_etal:08_mf}, halos are allowed to overlap so long as the
center of one halo is not contained within the radius of another
halo. Thus, the minimum separation of two halos with radii $R_1\ge
R_2$ is $R_1$, rather than the sum of the two radii, as done in
\cite{tinker_etal:05}.

\begin{figure}
\epsscale{1.0} 
\plotone{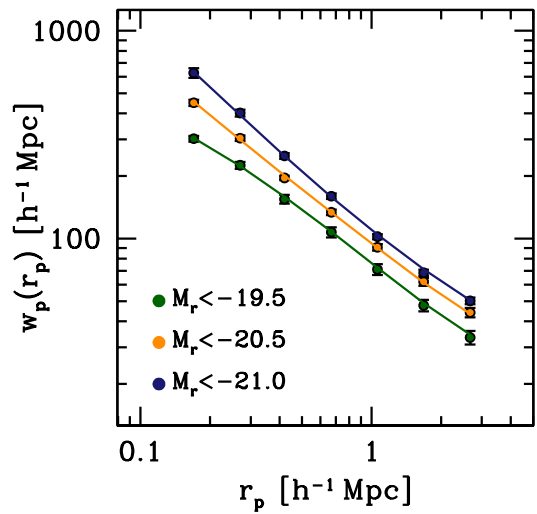}
%\vspace{-7.5cm}
\caption{ \label{bestfit_wp} The best-fit HOD model fits to $\wp$ for
  each $\mrz$ threshold. Uncertainties in the model fits are shown in
  Figure \ref{wpdiff}.}
\end{figure}

\subsection{Calculating $M/N$ from the HOD}

Measuring the galaxy content within a cluster sample stacked on
richness is not identical to measuring the mean number of galaxies
within the virial radii of halos of mass $M$. First, due to scatter in
the mass-richness relation, a stack of clusters with the same $\N$
contains a sample of halos of varying mass. Second, each richness bin
stacks clusters from a range of richness values (e.g., the most
massive richness bin stacks all clusters with $71\le\N\le220$). The
halo radius for each richness bin, $\RN$, is determined such that the
mean overdensity of all the halos in the bin is 200$\rhocrit$. The
true radii of individual halos in the bin, $R_M$ are not identical due
to scatter; some halos will be smaller the $\RN$ while some will be
larger. Both the scatter and the fixed aperture are accounted for when
we calculate $M/N$ within our HOD models.

For each bin of $\N$ values, the expected value of $M/N$ for a given
model is computed by

\begin{equation}
\label{e.m2n_theory}
M/N = \frac{\sum_{\N}\int dM n_h(M) P(\N|M) M  f_{\rm h}(\RN|M)}{\sum_{\N}\int dM n_h(M) P(\N|M) \nsat f_{\rm gal}(\RN|M)},
\end{equation}

%\begin{equation}
%\label{e.m2n_theory}
%M/N = \frac{\int dM n_h(M|\Delta_m,z) P(\N|M) \nsat f_{\rm gal}(R_M, \RN)}{\int dM %n_h(M|\Delta_m,z) P(\N|M) M  f_{\rm h}(R_M, \RN)},
%\end{equation}

\noindent where $n_h(M)$ is the halo mass function given by
\cite{tinker_etal:08_mf}\footnote{The \cite{tinker_etal:08_mf} mass
  function is universal at a given redshift for constant values of
  $\Delta_m$, the overdensity relative to the {\it mean} density in
  the universe. The maxBCG observations of mass and $N_{\rm gal}$ are
  made at 200 times the critical density ($\Delta_c$) assuming a value
  of $\om=0.27$ and $z=0.25$, which translates to $\Delta_m=508$.},
$P(\N|M)$ is the probability of a cluster with mass $M$ having
richness $\N$ (\citealt{rozo_etal:09_scatter}), and $f$ is the
aperture correction factor for both the mass and the number of
galaxies, given the fact that the radius of a halo of mass $M$, $R_M$,
may be different from the fixed aperture $\RN$ for that richness
bin. The integrals in the numerator and denominator are evaluated at
each value of $\N$ in the richness bin and summed together. Note
however, that $\RN$ is the same value within a richness bin; see the
values listed in Table 2.  The scatter in richness at fixed halo mass
is assumed to be a lognormal of the form

\begin{equation}
\label{e.scatter}
P(\N|M)d\N = \frac{1}{\sqrt{2\pi}\sigma_{R}}\exp\left[\frac{-(\ln\N-\mu_R)^2}{2\sigma_R^2}\right]\frac{d\N}{\N}.
\end{equation}

\noindent The mean of this distribution is

\begin{equation}
\label{e.mass_richness}
\mu_R = B_R + A_R\ln \left(M/M_{\rm pivot}\right),
\end{equation}

\noindent where $M_{\rm pivot}=2.06\times 10^{13}\times 0.7$ \hmsol,
and the factor of 0.7 is to change the value listed in
\cite{rozo_etal:09_scatter} from units of M$_\odot$ to \hmsol. We will
discuss the values and errors on the mass-richness scatter in the
following section.

The aperture correction factor is small but non-negligible. For a halo
with mass $M$, corresponding to radius $R_M$, and concentration
$c(M)$, the mass enclosed at any radius $R$ is calculated by

\begin{equation}
\label{e.f}
f_{\rm h}(R|M) = \frac{1}{M}4\pi\rho_sR^3 y^3\left[\ln(1+1/y)-(1+y)^{-1}\right]
\end{equation}

\noindent (c.f., appendix A in \citealt{hu_kravtsov:03}), where
$y\equiv r_s/R$ and $r_s\equiv R_M/c(M)$). Equation (\ref{e.f}) assumes
an NFW form for the halo density profile. The parameter $\rho_s$ is a
normalization parameter such that $f_{\rm h}(R_M|M)=1$. Because we
assume that satellite galaxies also follow an NFW profile, we can use
the same scaling relation to determine the number of satellite
galaxies within an aperture $R$ for a halo of mass $M$.

\begin{equation}
\label{e.fg}
f_{\rm gal}(R|M) = \frac{1}{\nsat}4\pi\rho_s\RN^3 y^3\left[\ln(1+1/y)-(1+y)^{-1}\right]
\end{equation}

\noindent where $\rho_s$ is once again determined by requiring $f_{\rm
  gal}(R_M|M)=1$. As discussed above, the concentrations parameter for
satellite galaxies is defined separately from the dark matter.

\subsection{Probing Cosmology with $\wp$ and $M/N$}
\label{s.hod_demo}

To demonstrate the power of combining $\wp$ and $M/N$, Figure
\ref{hod_demo} shows both statistics for a series of cosmological
models. In all panels, the four curves represent four HOD models
applied to the DR7 clustering measurements of the $\mrz<-20.5$ sample,
each with different values of $\s8$. For demonstration purposes, all
other cosmological parameters are fixed, and we fix $\asat=1$. For
each cosmological model, a good fit to the data can be found,
demonstrating that cosmology and bias are degenerate for this single
statistic. However, as the halo population changes with $\s8$, the
halo occupation must also change in a compensatory fashion in order to
obtain the same two-point clustering. In Figure \ref{hod_demo}c, $M/N$
is shown for each of the best-fit HOD models. The low-amplitude
cosmology ($\s8=0.7$) requires a larger weighting of high-mass (and
thus highly-biased) halos in order to match the observed amplitude of
the galaxy clustering. This drives $M/N$ lower relative to the other
cosmologies. The theoretical predictions also depend on the value of
$\om$; if the matter density increases, then the mass of each halo
increases proportionately, but the number of galaxies per halo remains
fixed. Thus $\om$ and $\s8$ will be degenerate in this analysis much
as they are in cluster abundance constraints

The comparison here is not one-to-one, given that the data are
convolved with the mass-richness scatter and that we have not
marginalized over the many free parameters in our model. Nonetheless,
this plot demonstrates that the degeneracy of cosmology and bias with
respect to the two-point correlation function can be broken by
empirical measurements of the HOD in the form of $M/N$.

\begin{figure}
\epsscale{1.5} 
\plotone{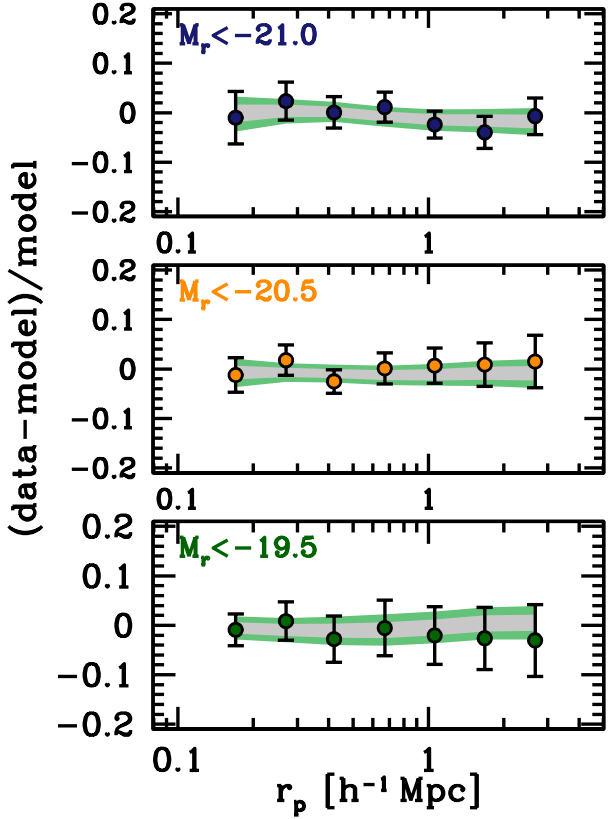}
%\vspace{-7.5cm}
\caption{ \label{wpdiff} Each panel shows the fraction difference
  between the measurements of $\wp$ and the best-fit model for each
  $\mrz$ threshold. The contours indicate the range of models from the
  MCMC chain. The inner contours include 68\% of all elements in the
  chain, while the outer contours bracket 95\% of the elements in the
  chain. The data points are highly correlated, thus we caution the
  reader against $\chi$-by-eye. Note that all models are
  simultaneously fit to $M/N$ as well as $\wp$.}
\end{figure}

\begin{figure}
\epsscale{1.5} 
\plotone{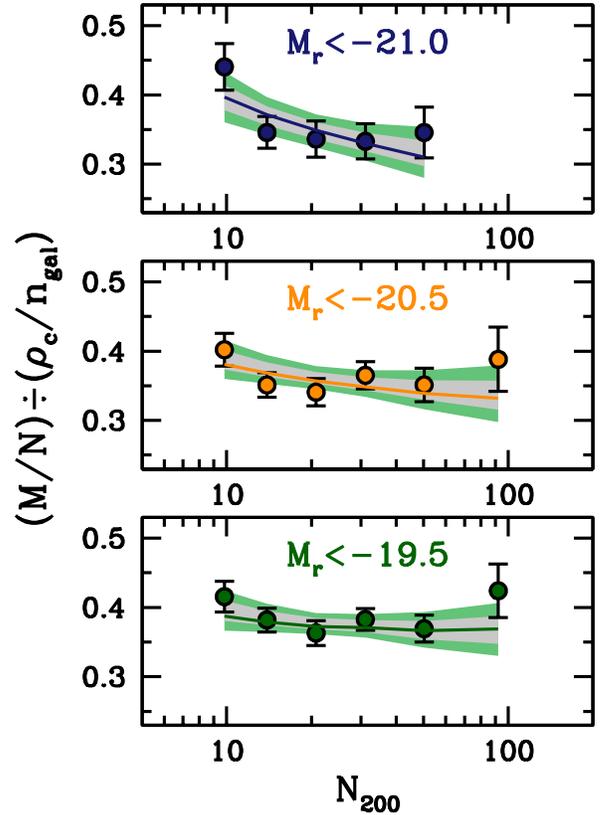}
%\vspace{-7.5cm}
\caption{ \label{bestfit_m2n} Each panel shows the $M/N$ measurements,
  normalized by $\rhocrit/\ngavg$, for each threshold sample. The
  solid curve in each panel shows the best-fit HOD+cosmological model. The inner
  and outer shaded regions bracket the 68\% and 95\% ranges of models
  in the MCMC chain, respectively. Note that all models are
  simultaneously fit to $\wp$ as well as $M/N$.}
\end{figure}

\begin{figure}[t!]
\epsscale{1.2} 
\plotone{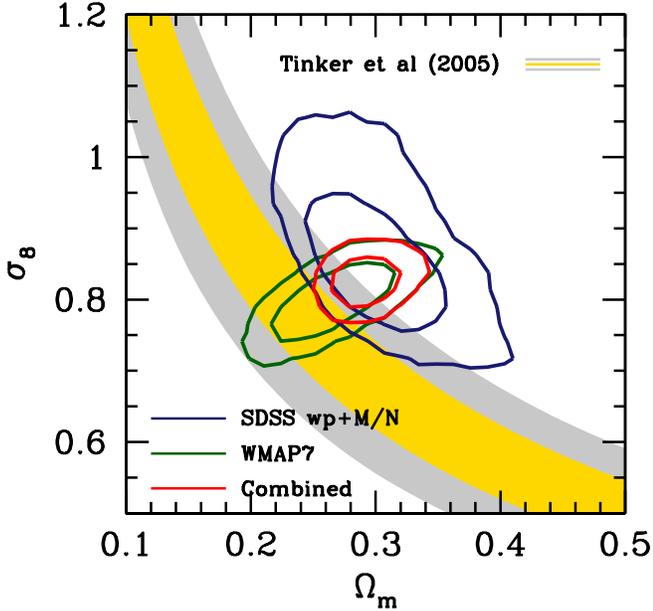}
%\vspace{-7.5cm}
\caption{ \label{banana1} Constraints in the $\om$-$\s8$ plane,
  marginalizing over all other parameters and applying the priors
  listed in Table 3. The blue contours show 68\% and 95\% constraints
  from the $M/N$ results. The yellow and grey shaded region indicates
  the $1\sigma$ and $2\sigma$ constraints, respectively, from
  \cite{tinker_etal:05}. The green contour shows the constraints from
  WMAP7 (CMB alone, assuming a flat-\lcdm\ model;
  \citealt{komatsu_etal:10}). The red contours show the combined
  constraints from $M/N$ and WMAP7.}
\end{figure}

\begin{figure*}
\epsscale{1.0} 
\plotone{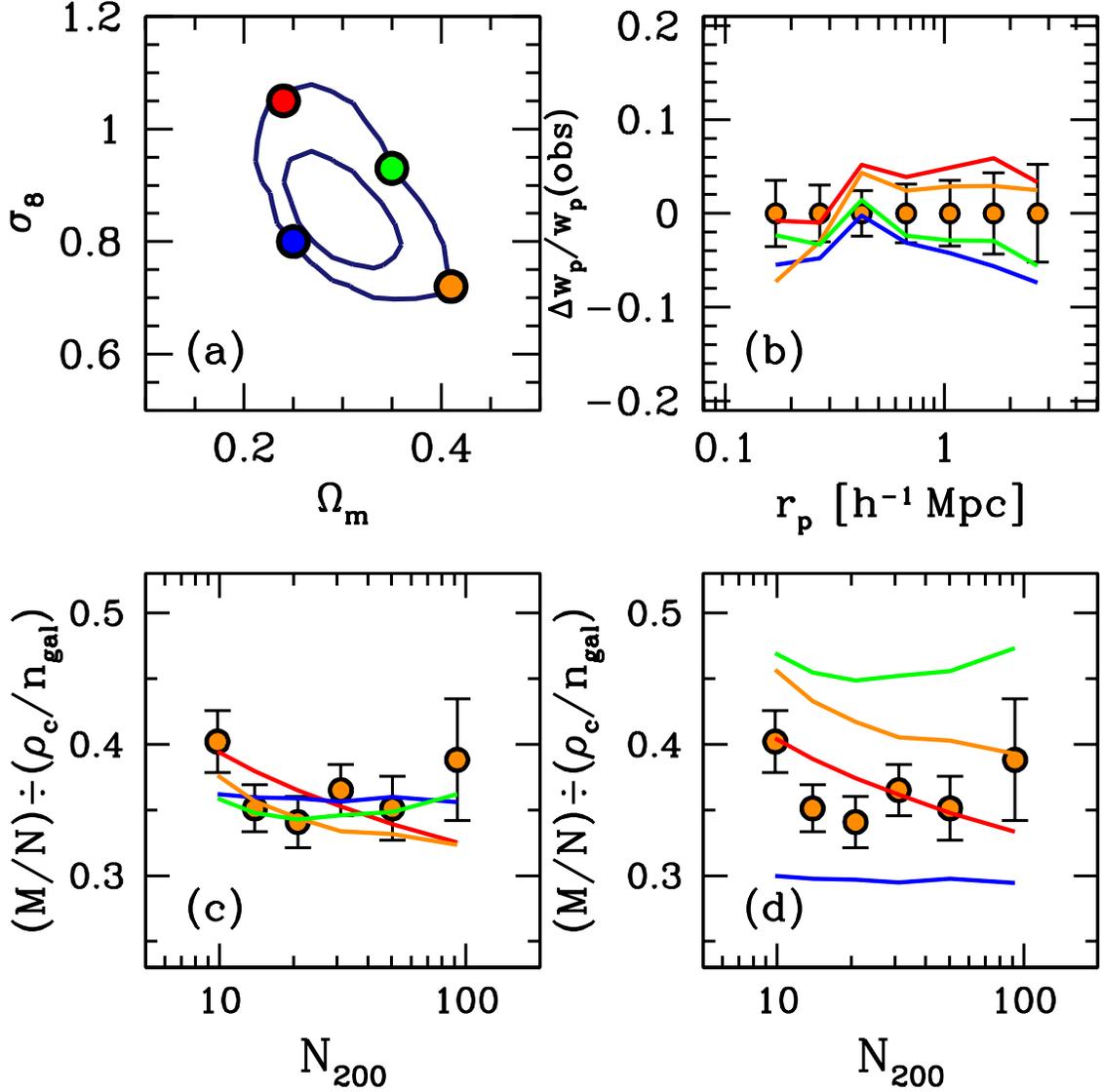}
%\vspace{-7.5cm}
\caption{ \label{constraints} Panel (a): The circles indicate the
  location of four models in the $\om$-$\s8$ plane chosen from the
  MCMC chain. Each model is the best fitting model for that region of
  cosmological parameter space. Panel (b): Comparison of each model to
  the $\wp$ data. The $y$-axis is
  $(w_p^{HOD}-w_p^{obs})/w_p^{obs}$. The points with error bars are
  the $\mrz<-19.5$ measurements. Panel (c): Comparison of each model
  to the $M/N$ measurements, again for the $\mrz<-20.5$ sample. The
  fits appear good because $\fsys$ is allowed to vary from
  model-to-model; each model requires an unlikely value of $\fsys$ to
  fit the data. Panel (d): The predictions of those same models when
  $\fsys=1.13$, the value in the minimum $\chi^2$ fit.}
\end{figure*}

\begin{figure*}
\epsscale{1.2} 
\plotone{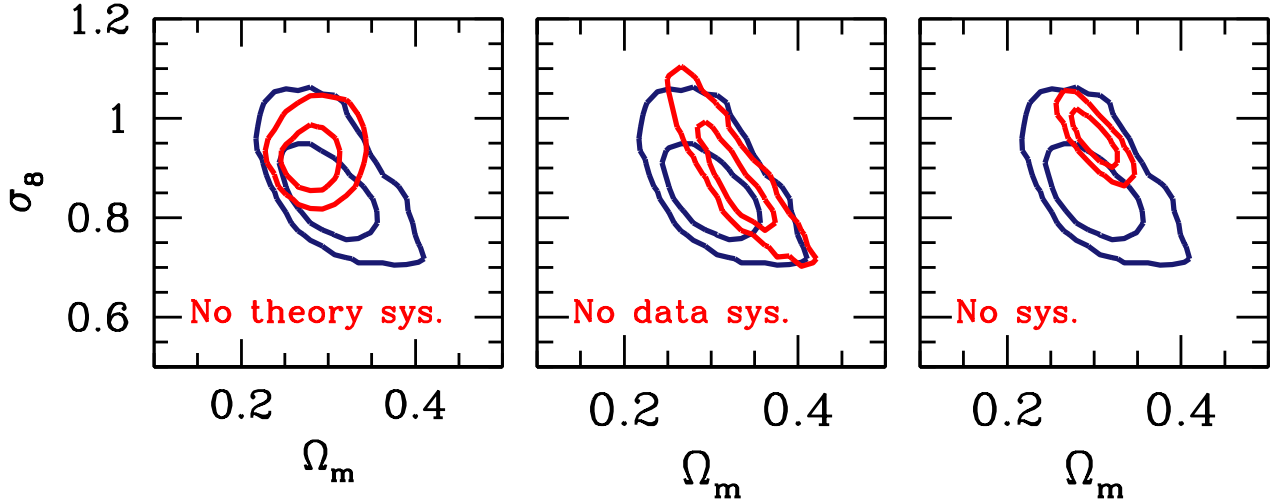}
\vspace{-11cm}
\caption{ \label{systematics_banana} The effect of systematic
  uncertainties on our cosmological parameter constraints. In all
  panels, the blue contour show our fiducial results from Figure
  \ref{banana1}. {\it Left panel:} The red contour indicates the
  parameter constraints when reducing the uncertainty in the dark
  matter halo statistics by a factor of 5. Specifically, we reduce the
  priors on $\en$, $\eb$, and $\en$ to 0.01, 0.01, and 0.03,
  respectively. The dominant uncertainty in this panel is from the
  large-scale halo bias. {\it Middle Panel:} The effect of systematic
  uncertainties in our measurements. The red contour shows the
  parameter constraints when reducing the prior on $\fsys$ from 0.227
  to 0.02. Note that $\fsys$ represents a combined effect of (in
  decreasing order of importance) uncertainties in the $z=0.25$
  luminosity function, evolution of the HOD from $z=0.1$ to $z=0.25$,
  and the overall calibration of weak lensing masses. {\it Right
    panel:} The red contour shows the effect of reducing {\it both}
  the theoretical and measurements uncertainties from the left two
  panels. }
\end{figure*}

\begin{figure*}
\epsscale{1.2} 
\plotone{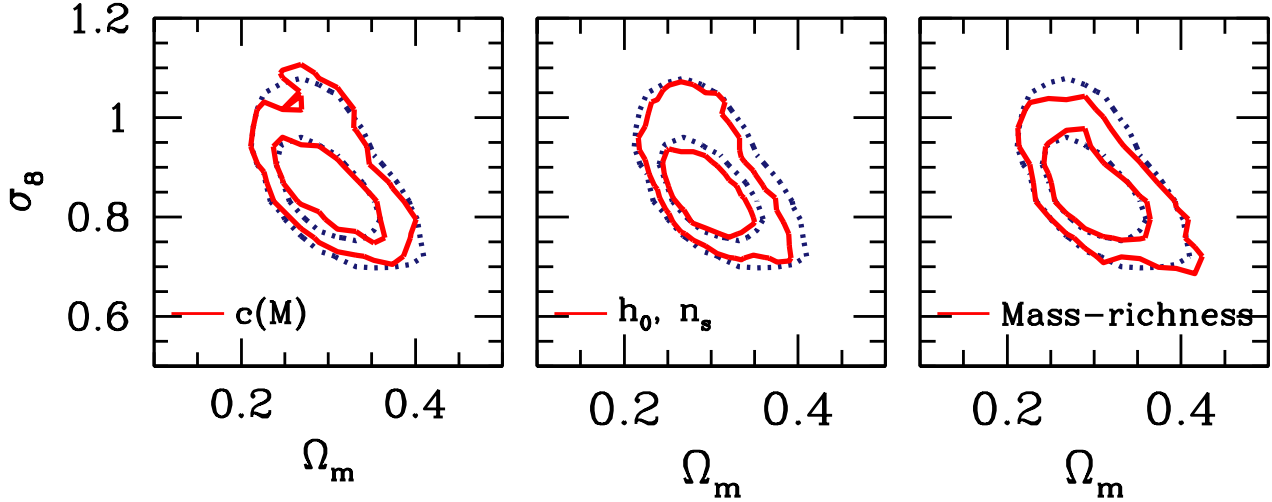}
\vspace{-11cm}
\caption{ \label{systematics_banana2} The effect of our priors on
  cosmological constraints. In each panel, the dotted contour shows
  the fiducial results from Figure \ref{banana1}. {\it Left Panel:}
  Red solid contours show parameter constraints when changing the
  prior on $\fcon$ from flat to a 10\% Gaussian prior. {\it Middle
    Panel:} Red solid contours show parameter constraints when
  tightening our priors on $h_0$ and $n_s$ from 0.05 and 0.03 to 0.02
  and 0.01. The minimal change in the constraints shows that our
  results are insensitive to the allowed range in the shape of the
  matter power spectrum. {\it Right Panel:} Red solid contours show
  the parameter constraints obtained when tightening the priors on the
  mass-richness relation: $B_R$, $A_R$, and $\sigma_R$. Our fiducial
  errors on these quantities are 0.09, 0.024, and 0.07,
  respectively. The solid curves show results for uncertainties of
  0.01, 0.005, and 0.01. These results indicate that the $M/N$ method
  is largely insensitive to scatter in the mass-richness relation. }
\end{figure*}

\begin{figure*}
\epsscale{1.0} 
\plotone{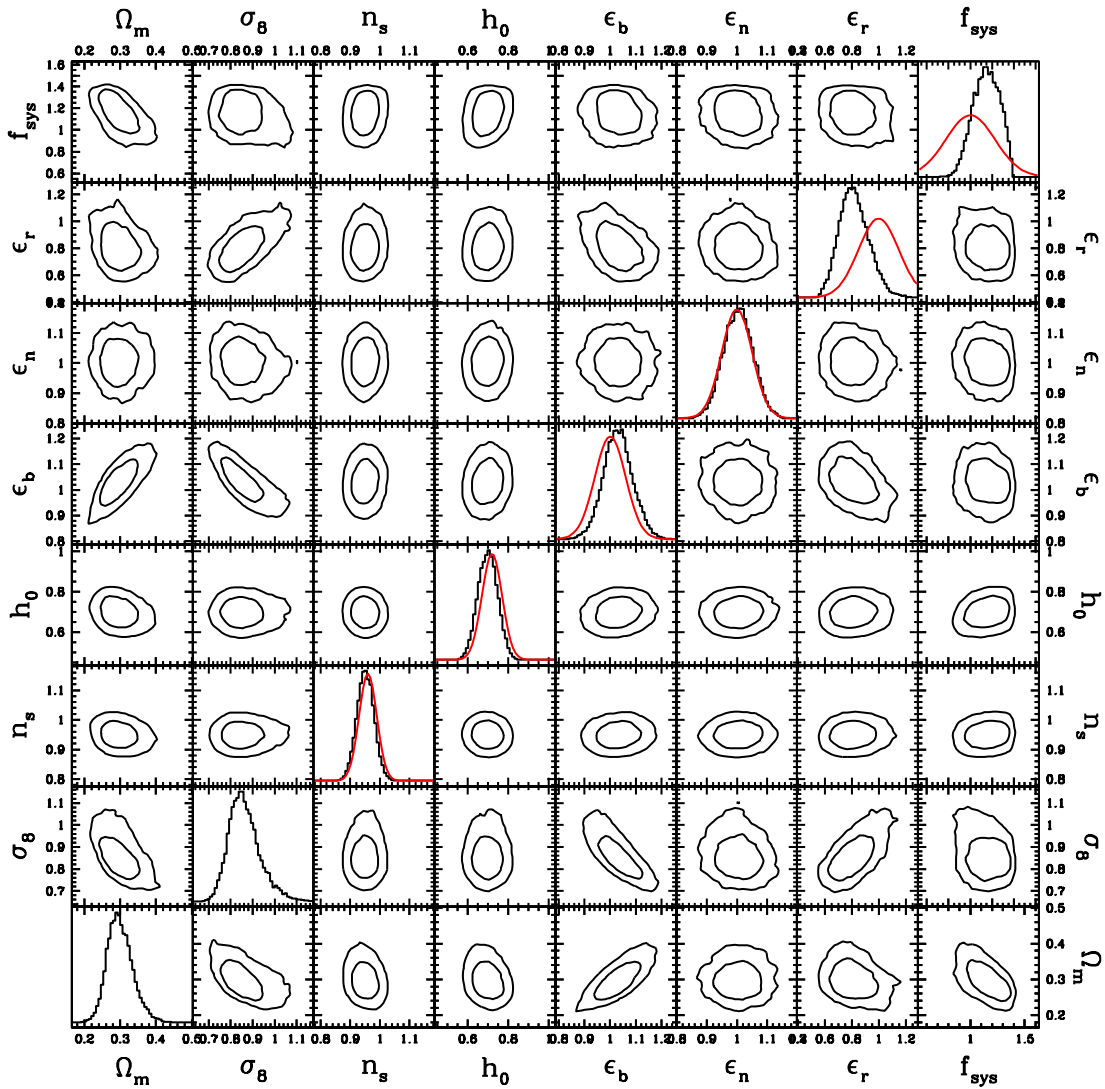}
%\vspace{-7.5cm}
\caption{ \label{grid_sys} Each panel shows 68\% and 95\% constraints
  for our cosmological parameters and the marginalization parameters
  that encompass our systematic uncertainties. The histograms in
  diagonal panels shows the distribution of each parameter from the
  MCMC chain. The red curves in the diagonal panels show the priors on
  each parameter. Parameter definitions are given in Table 3.}
\end{figure*}

\begin{figure}[t!]
\epsscale{1.0} 
\plotone{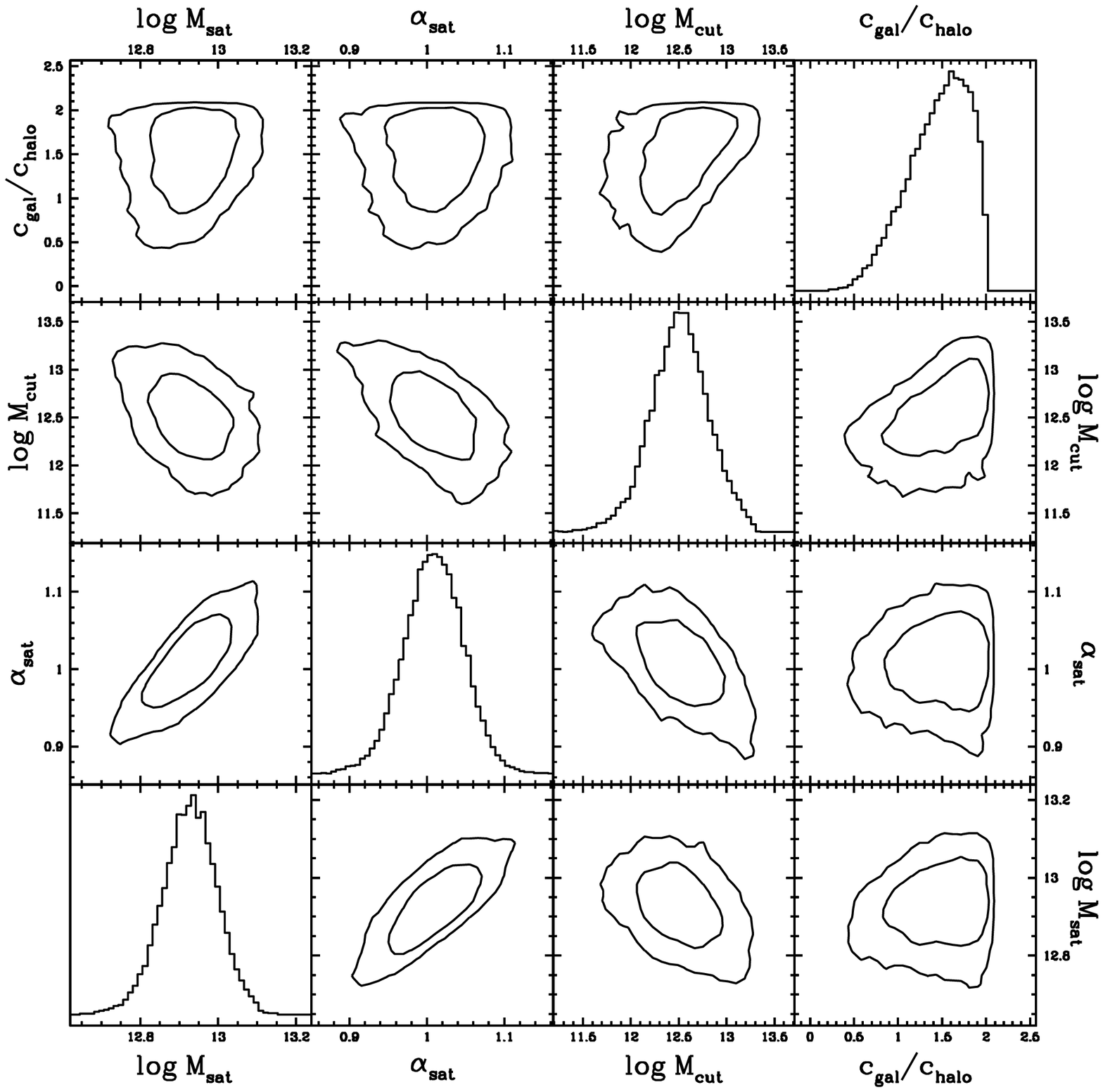}
%\vspace{-7.5cm}
\caption{ \label{grid_19.5} HOD parameter constraints for the
  $\mrz<-19.5$ sample. The diagonal panels show the distribution of
  values from the MCMC chain. Note that $\slogm$ is not a free
  parameter in this sample.}
\end{figure}

\begin{figure}[t!]
\epsscale{1.0} 
\plotone{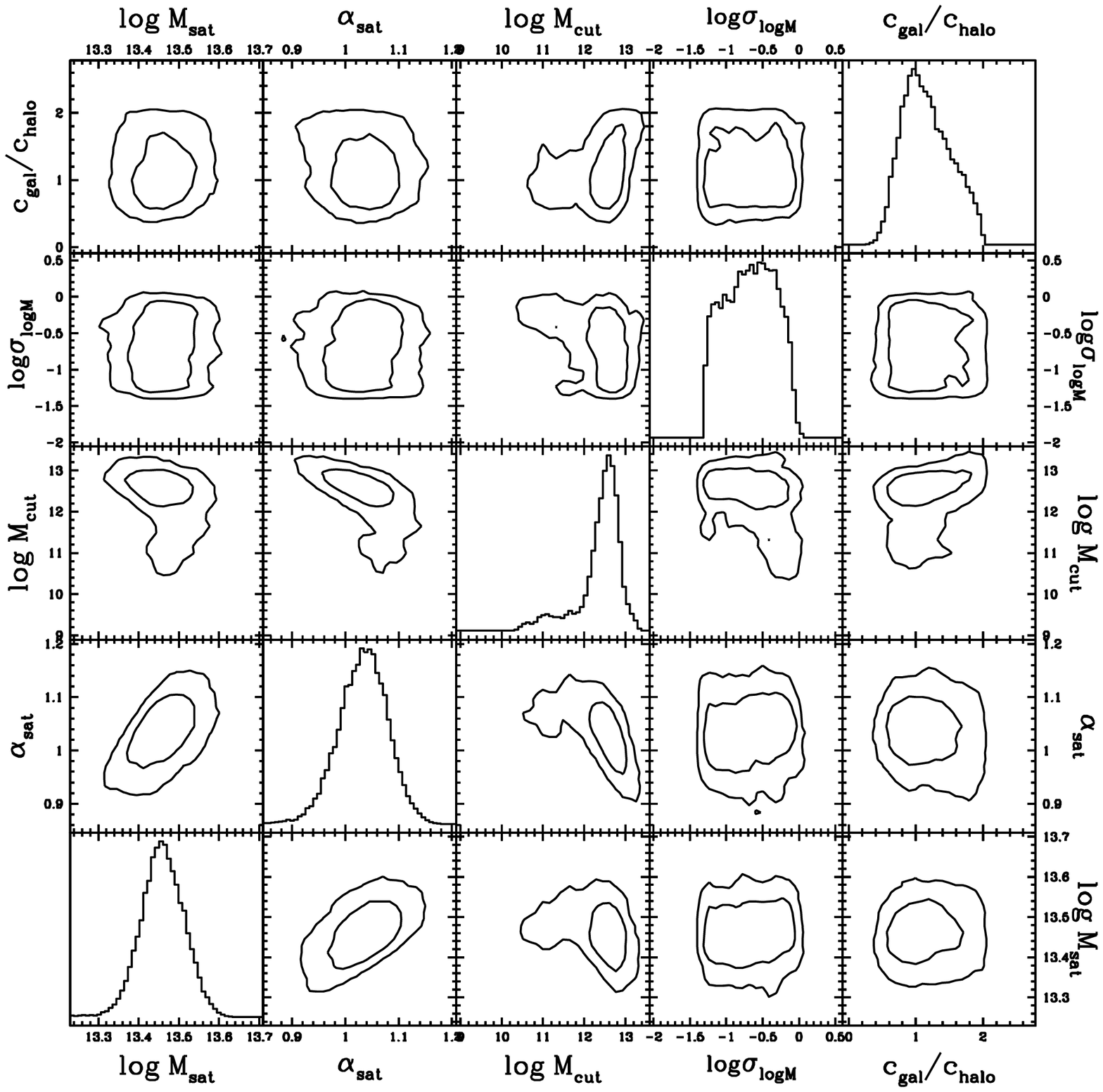}
%\vspace{-7.5cm}
\caption{ \label{grid_20.5} HOD parameter constraints for the
  $\mrz<-20.5$ sample. The diagonal panels show the distribution of
  values from the MCMC chain. }
\end{figure}

\section{Systematic Errors}
\label{s.systematics}

Here we detail the many systematic errors that we consider in our
analysis.  Some are associated with theoretical uncertainties, others
with observational uncertainties.

{\it Uncertainty in the halo mass function}: We use the
simulation-calibrated halo mass function of \cite{tinker_etal:08_mf}
in all theoretical calculations. This mass
function is calibrated on spherical overdensity halos, where halo mass
is defined by the mass within a spherical aperture, the same manner as
the cluster mass in the maxBCG weak-lensing mass estimates. The
\citeauthor{tinker_etal:08_mf} mass function also takes into account
redshift evolution of the mass function, which is 10-20\% between
$z=0$ and $z=0.25$. Most of the simulations analyzed in
\cite{tinker_etal:08_mf} are realizations of a single, flat-\lcdm\
cosmology representative of first-year WMAP results
(\citealt{spergel_etal:03}). For this cosmology, the errors on the
mass function are less than 1\% up to nearly $10^{15}$ \hmsol. For
variants around this cosmology, the simulations produce a 5\%
scatter. We therefore incorporate a 5\% Gaussian uncertainty on the
amplitude of the halo mass function, introducing a marginalization
parameter $\en=1.0\pm 0.05$, which represents the amplitude of the
halo mass function relative to the \cite{tinker_etal:08_mf} fit.  We
describe the incorporation of this and subsequent uncertainties at the
end of this section.

{\it Uncertainty in the large-scale halo bias relation}: The
\citeauthor{tinker_etal:08_mf} mass function is coupled to the bias
functions of \cite{tinker_etal:10_bias}. These bias functions are
calibrated on the same N-body simulations as the mass function and
with the same halo definitions. Thus, the abundance and clustering of
halos are always calculated using a self-consistent halo
definition. \cite{tinker_etal:10_bias} find a 6\% scatter in bias
among the simulations. To be conservative, we implement a 6\%
Gaussian error on the amplitude of the halo bias relation. As with the
mass function, we delineate the prior on the bias amplitude as $\eb$,
with a mean of 1.0 and an uncertainty 0.06.

{\it Uncertainty in the scale-dependence of the halo bias}: At scales
of $r\lesssim 5$ \hmpc, halo clustering deviates from a
scale-independent factor of the matter clustering. This scale
dependence is minimized when the non-linear matter clustering (rather
than the linear) is used to define the halo bias. We assume that the
shape of the scale-dependence follows that of \cite{tinker_etal:05}
but allow the magnitude to differ. We include a 15\% Gaussian error in
the deviation from constant bias, i.e.,

\begin{equation}
\label{e.scale_error}
  \delta b = b(r) - b_0,
\end{equation}

\noindent where $b(r)$ is the scale-dependent bias of
\cite{tinker_etal:05} (with a modification for very small scale halo
clustering given in Appendix A) and $b_0$ is the large-scale bias of
\cite{tinker_etal:10_bias}. The \cite{tinker_etal:05} scale-dependent
bias, $b(r)/b_0$, asymptotes to unity at $r\gtrsim 5$
\hmpc. Parameterizing the uncertainty in the scale dependence through
eq. (\ref{e.scale_error}) ensures that bias becomes linear at large
scales. We refer to this prior as $\er$, with an uncertainty of 0.15.

{\it Uncertainty in the parameters of the mass-observable relation}:
As described above, \cite{rozo_etal:09_scatter} describe the relation
between halo mass and observed cluster richness from the maxBCG
algorithm as a power law with parameters $\alpha_R$ and $\beta_R$. The
scatter of mass at fixed richness is modeled as a lognormal with
variance $\sigma_R$. Using X-ray data from the ROSAT All-Sky Survey,
\cite{rozo_etal:10} both calibrate this relation and determine
the uncertainties in each parameter. These are found to be

\begin{eqnarray}
A_R = 0.750 \pm 0.024 \nonumber \\
B_R = -1.09 \pm 0.09 \nonumber \\
\sigma_R = 0.35 \pm 0.07. \nonumber 
\end{eqnarray}

\noindent These three parameters are also added to the chain with
the uncertainties listed above.

{\it Uncertainty in the evolution of the HOD:} Due to the redshift
baseline of our two galaxy samples, $z=0.1$ and $z=0.25$, we must
account for possible evolution in the HOD over this narrow
range. Analysis by \cite{zheng_etal:07} over a much larger baseline,
from $z=0.1$ to $z=1$ using the clustering of DEEP2 galaxies,
demonstrates that the $\msat/\mmin$ ratio evolves from 18 to 16
(independent of luminosity). \cite{abbas_etal:10} see even less
evolution from $z=0.4$ to $z=1.1$, but using the smaller zCOSMOS data
set.

We have investigated this issue theoretically in two ways: through
analysis of the {\it Millennium} semi-analytic galaxy catalogs of
\cite{bower_etal:06}, and through the high-resolution N-body
simulation Bolshoi (\citealt{klypin_etal:10}). Recall that we are
analyzing samples constructed to have the same space densities at
their respective redshifts, not the same magnitudes. This minimizes
the effect of the evolution of the galaxy population.  In the
semi-analytic {\it Millennium} model, the HOD at fixed $\ngavg$ shows
a moderate increase of 5-10\% in $\nsat$ at a given $M$ from $z=0.1$
to $z=0.25$.

In the Bolshoi simulation, we investigate the the evolution of the
halo occupation of {\it subhalos}. The connection between substructure
and galaxies is well-established (see, e.g.,
\citealt{kravtsov_etal:04, conroy_etal:06, moster_etal:09,
  behroozi_etal:10} for some examples), thus this method can be used
to construct galaxy catalogs without attempting to model galaxy
formation physics. At fixed space density\footnote{Here we mean
  density of all halos and subhalos above a given maximum circular
  velocity. For subhalos, the value of $V_{\rm max}$ at the time of
  accretion is used to connect to galaxy luminosity.}, the halo
occupation of subhalos shows $\lesssim 10\%$ evolution over our
redshift range (Reddick et al, in preparation), varying slightly with
galaxy number density. \cite{simha_etal:10} find that subhalo
occupation statistics track the the statistics of galaxy occupations
in smoothed particle hydrodynamics (SPH) simulations fairly
accurately, with the largest deviations arising because a small but
not negligible fraction of SPH galaxies suffer dramatic stellar mass
loss after entering high mass halos, thus ending up less massive than
predicted. 

Based on these results, we incorporate HOD evolution uncertainty as a
10\% Gaussian multiplicative error on $M/N$, centered on no evolution.

{\it Uncertainty in the luminosity function at $z=0.25$}: As odd as it
may seem, there is no reliable measurement of the $^{0.25}M_r$
luminosity function. As discussed above and shown in Figure
\ref{lumfunc_evolution}, we infer this function by the method of
\cite{blanton:06}. Although the redshift difference between the Main
sample and the maxBCG sample is relatively small, the amplitude of
$M/N$ is sensitive to the details of our luminosity function
estimate. The key quantities are the $^{0.25}M_r$ magnitudes that
yield the same number densities as the $z=0.1$ samples. A 0.1 mag
error in these values yields a $\sim 20\%$ error in $M/N$. Given that
the overall evolution of the luminosity function is 0.2 to 0.3
magnitudes, we use 0.1 magnitudes as a conservative error estimate on
our estimate of the $z=0.25$ luminosity function. As with the
evolution of the HOD, evolution in the luminosity function shifts the
amplitude of $M/N$, thus we include an additional 20\% Gaussian
multiplicative error on our theoretical calculation of $M/N$ for a
given model. The quantitative incorporation of this uncertainty with
HOD evolution will be discussed at the end of this section.

{\it Uncertainty in the calibration of the weak lensing masses}:
\cite{rozo_etal:09_scatter} apply a correction factor of $1.18\pm
0.06$ to the maxBCG halo masses inferred through weak lensing in
\cite{sheldon_etal:09_data} and \cite{johnston_etal:07}. This factor
is to compensate for the fraction of sources with redshift errors
large enough to place their true redshift in front of the lensing
clusters, thereby diluting the shear signal. An error in halo mass
scale is partly compensated by an increase in the radius at which the
mean interior density is 200$\rhocrit$, which is the radius at which
$N_{\rm sat}$ is measured. When measuring $M/N$ for both the original
\cite{sheldon_etal:09_data} masses and the new corrected masses, we
find that the 18\% increase in the mass at each richness bin yields
only a 9\% increase in the $M/N$ ratio at each richness bin. The
remaining 6\% uncertainty in the halo masses therefore yields a 3\%
uncertainty in $M/N$. This error is correlated over all data points,
and it is included with the other data systematics listed above:
evolution of the HOD and the luminosity function.

{\it Miscentering of the maxBCG algorithm}: As with all optical
cluster finders, maxBCG is not 100\% pure and complete. Some fraction
of the time, the correct halo is identified but the wrong galaxy is
identified as the BCG. Additionally, chance projection of two clusters
can result in the two objects being merged in the resulting cluster
catalog. In the mass estimates of \cite{sheldon_etal:09_data}, these
effects are not accounted for. They are also not accounted for in the
measurements of the galaxy counts. Because satellite galaxies follow
roughly the same distribution as the dark matter within the cluster,
the effect on their ratio is significantly smaller than on either
quantity individually. \cite{rozo_etal:10} find that the fraction of
chance projections is small compared to the intrinsic scatter in the
mass-richness relation, thus they conclude that the effect of
projections on their constraints is negligible.

We quantify the effect of miscentering using mock galaxy catalogs on
which the maxBCG algorithm is run. The mock catalogs use the ADDGALS
(Adding Density Determined Galaxies to Lightcone Simulations)
algorithm to populate a dark matter simulation with galaxies
(\citealt{wechsler:04}; Wechsler et al preparation).  Galaxies
are assigned to dark matter particles based on their dark matter
densities, constrained to match the observed two-point galaxy
clustering statistics and luminosity function, and are assigned galaxy
colors based on the observed distribution of galaxy color at a given
galaxy density and luminosity.  Mocks created with this algorithm has
been used to test the maxBCG algorithm in several previous works
(e.g.~\citealt{koester_etal:07, johnston_etal:07, rozo_etal:10,
  gerdes_etal:10}).  Clusters in which the detected BCG is the same as
a central galaxy in a mock dark matter halo are deemed
``well-matched''. Clusters where this match could not be made
represent objects where the detected BCG is not a central galaxy. For
the well-matched sample we repeat all the steps described in Figure
\ref{maxbcg_data}: For each bin in richness, the mock
background-subtracted surface density of galaxies is measured and
inverted to obtain $n_{\rm gal}(r)$. This quantity is integrated out
to $R_{\rm halo}$, yielding $N_{\rm sat}$. The same process is
performed on the dark matter particles as well to mimic the effect of
miscentering on the weak lensing masses and the $M/N$ ratio. To obtain
$M_{\rm halo}$ and $R_{\rm halo}$ the dark matter density profile is
integrated out until $\Delta=200\rhocrit$. The results are shown in
Figure \ref{miscentering_correction} for all three luminosity
thresholds. At high richness, miscentering has a negligible effect on
$M/N$. At $N_{200}\lesssim 20$, $M/N$ for the well-matched sample is
larger than that of the full sample of clusters. This implies that the
effect of miscentering on mass is somewhat stronger than on galaxy
number. However, the effect of miscentering on $M/N$ is significantly
smaller than on $M$ by itself, as shown by the circles connected by
the dotted line.  Although there appears to be a monotonic effect with
luminosity at low richness, it is not clear if this trend is real or
noise. We take the mean of these three curves as the miscentering
correction, with the range between the curves as a $1\sigma$ Gaussian
error on the correction. This error is added in quadrature to the
statistical errors on $M/N$. It is listed in Table 2.

{\it Environmental dependence of the HOD}: It has been demonstrated
that the properties of halos at fixed mass depend on large-scale
environment, an effect dubbed `assembly bias' (see, e.g.,
\citealt{gao_white:06, wechsler_etal:06}). Semi-analytic models of
galaxy formation have predicted that this assembly bias could
propagate into the properties of galaxies within halos at fixed mass
(\citealt{croton_etal:07, zhu_etal:06}), although results from
hydrodynamic simulations do not support this
(\citealt{berlind_etal:03, yoo_etal:06}). In the standard HOD, $\navg$
depends only on $M$ and not on a second parameter, such as
$\delta$. If occupation depends on environment, then the large-scale
bias of a galaxy sample may differ from that calculated through the
standard HOD, biasing the results of the modeling. By modeling both
$\wp$ and the distribution of galaxy voids in the SDSS,
\cite{tinker_etal:08_voids} found that any assembly bias must have
less than 5\% effect on the large-scale clustering of
luminosity-defined samples for the luminosity range probed by SDSS
data. Furthermore, our exclusion of clustering data at $r_p>3$ \hmpc\
attenuates any impact assembly bias may have on our HOD fits. We
conclude, given the small effect of assembly bias and the exclusion of
large-scale data, that our 10\% prior on the evolution of the HOD
subsumes any error accrued due to assembly bias.

{\it Combination:} The uncertainties in evolution of the HOD, the
$z=0.25$ luminosity function, and the calibration of the weak lensing
masses all have the effect of systematically shifting the $M/N$
measurements up or down proportionately for all richness bins. These
three systematic errors are uncorrelated, thus we include a
$\sqrt{10^2+20^2+3^2}=22.6\%$ systematic error on all the $M/N$
data. To incorporate this uncertainty in the MCMC analysis, we add a
`systematic bias parameter' $\fsys$ to the chain. For each element in
the chain, the HOD prediction for $M/N$ is multiplied by $\fsys$. We
enforce a Gaussian prior of $1.0 \pm 0.226$ on $\fsys$.

\begin{figure}[t!]
\epsscale{1.0} 
\plotone{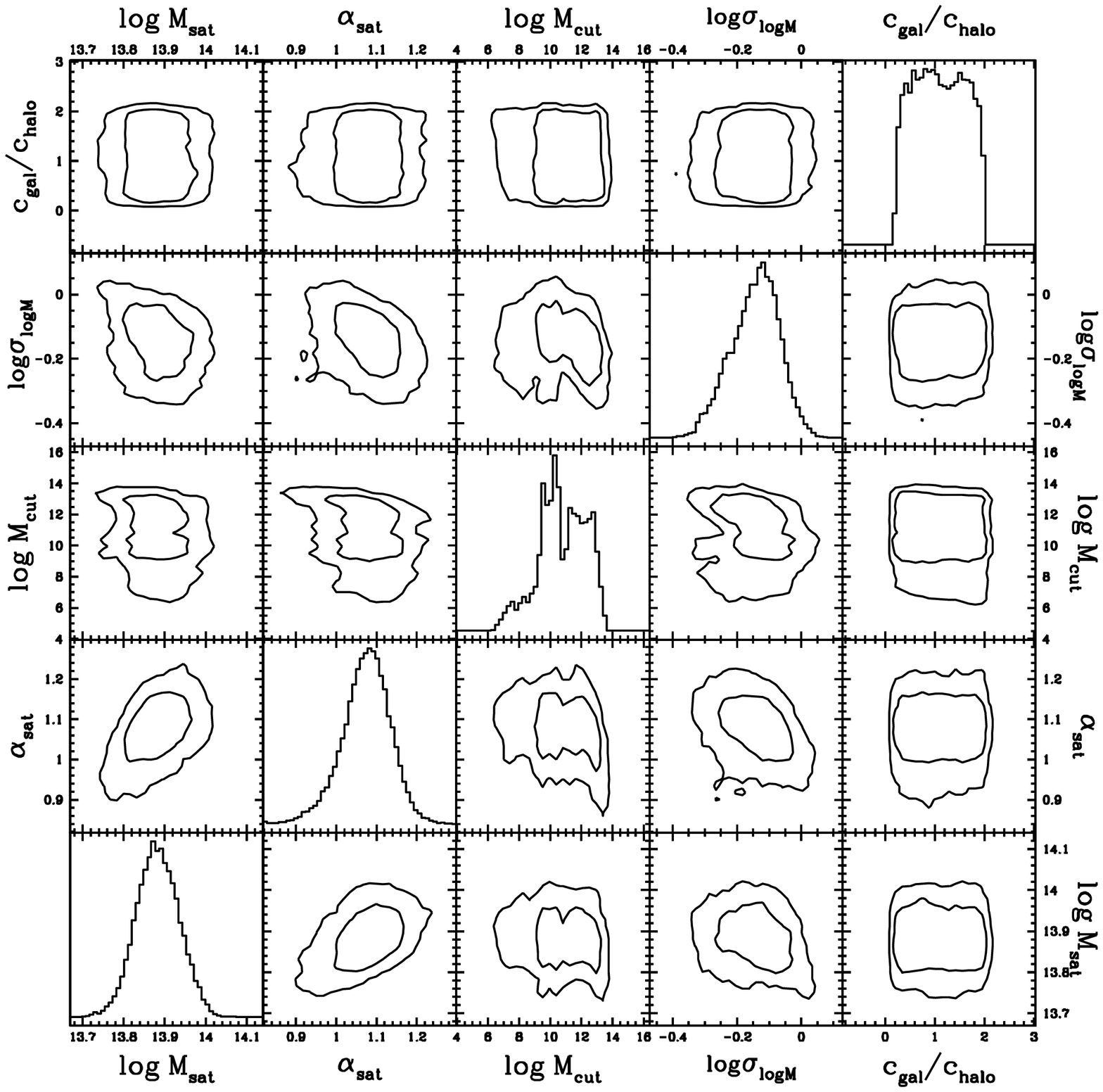}
%\vspace{-7.5cm}
\caption{ \label{grid_21.0} HOD parameter constraints for the
  $\mrz<-21.0$ sample. The diagonal panels show the distribution of
  values from the MCMC chain.}
\end{figure}

%%%%%%%%%%%%%%%%%%%%%%%%%%%%%%%%%
% Table 3
%%%%%%%%%%%%%%%%%%%%%%%%%%%%%%%%%
\begin{deluxetable*}{cccl}
  \tablecolumns{4} 
  \tablewidth{35pc} 
  \tablecaption{List of Parameters}
  \tablehead{\colhead{Name} & \colhead{Type} & \colhead{Value} & \colhead{Meaning}}

\startdata

$\mrz$ & observational & --- & SDSS $r$-band absolute magnitude $k+e$ corrected to $z=0.1$.\\
$^{0.25}M_r$ & observational & --- & SDSS $r$-band absolute magnitude $k+e$ corrected to $z=0.25$.\\
$\om$ & cosmological & no prior & matter density relative to the critical density\\
$\s8$ & cosmological & no prior & amplitude of linear matter fluctuations on the 8 \hmpc\ scale\\
$h_0$ & cosmological & $0.72\pm 0.05$ & Hubble constant in units of 100\,km/s/Mpc\\
$n_s$ & cosmological & $0.96\pm 0.03$ & spectral index of primordial fluctuations\\
$\mmin$ & HOD & no prior & central occupation function; see eq. (\ref{e.ncen}). \\
$\slogm$ & HOD & [0.05,1.6] &  halo mass-galaxy luminosity scatter; see eq. (\ref{e.ncen}). \\
$\msat$ & HOD & no prior & amplitude of the satellite occupation function; see eq. (\ref{e.nsat}). \\
$\mcut$ & HOD & no prior & cutoff of the satellite occupation function; see eq. (\ref{e.nsat}). \\
$\asat$ & HOD & no prior & power-law slope of the satellite occupation function; see eq. (\ref{e.nsat}). \\
$\fcon$ & HOD & [0.2,2] & satellite galaxy density profile: $\fcon\equiv c_{\rm gal}/c_{\rm halo}$\\
$\en$ & systematic & $1.00\pm 0.05$ & uncertainty in amplitude of halo mass function \\
$\eb$ & systematic & $1.00 \pm 0.06$ & uncertainty in amplitude of halo bias relation \\
$\er$ & systematic & $1.00 \pm 0.15$ & uncertainty in amplitude of scale-dependent bias \\
$\fsys$ & systematic & $1.00 \pm 0.22$ & uncertainty in the measured amplitude of $M/N$\\
$B_R$ & maxBCG & $-1.09\pm 0.09$ & amplitude of mass-richness relation; see eq. (\ref{e.mass_richness}) \\
$A_R$ & maxBCG & $0.750 \pm 0.024$ & slope of mass-richness relation; see eq. (\ref{e.mass_richness}) \\
$\sigma_R$ & maxBCG & $0.35\pm 0.07$ & scatter in mass-richness relation; see eq. (\ref{e.scatter})\\
$N_{200}$ & maxBCG (observed)& --- & richness of maxBCG clusters \\
$\RN$ & maxBCG (observed) &--- & cluster radius at which $\Delta=200\rhocrit$ in a given richness bin\\
$N_{\rm sat}$ & maxBCG (observed) & --- & true number of satellite galaxies in maxBCG clusters\\

\enddata \tablecomments{The HOD parameters, $\mmin$, $\msat$,
  $\asat$, $\mcut$, $\fcon$, are constrained separately for all three
  clustering samples. Thus there are three values of each parameter
  that included in the analysis. The HOD parameters $\slogm$ is free
  for the brighter two samples and fixed at $\slogm=0.2$ for the faint
  sample. $\mmin$ for each sample is set by the number density of each
  sample once the other HOD parameters are specified. Thus there are
  14 free HOD parameters total in this analysis. }
\end{deluxetable*}
%%%%%%%%%%%%%%%%%%%%%%%%%%%%%%%%%
% END Table 3
%%%%%%%%%%%%%%%%%%%%%%%%%%%%%%%%%

\section{Results}

\subsection{Fitting the Data}

We determine constraints on all free parameters through the Monte
Carlo Markov Chain (MCMC) method. For each element in the chain, we
determine the $\chi^2$ for the model in the following manner. First,
$\chi^2$ is calculated independently for all three sets of $\wp$
measurements using the full covariance matrix for each sample. Next,
the $\chi^2$ for the 17 $M/N$ data points is calculated using
eq. (\ref{e.m2n_theory}) and the covariance matrix. The total $\chi^2$
for each model is the sum of the two. To marginalize over the
systematic uncertainty in the mass function, the parameter $\en$ is a
free parameter within the chain. For each element $i$ in the chain,
the mass function for that model is $n_{h,i}(M) = n_h(M)\times
\en$. While running the chain, we adopt a 5\% Gaussian prior on this
parameter. We marginalize over all other sources of systematic
uncertainty in the same manner, i.e., we repeat this procedure for
$\eb, \er, \fsys$, and the parameters of the mass-richness relation.

Our model has a total of 21 parameters and priors. A list can be found
in Table 3. There are 14 HOD parameters; each galaxy sample is modeled
with an HOD of 4 or 5 parameters (we exclude $\slogm$ from the
$\mrz<-19.5$ sample). There are two free cosmological parameters,
$\om$ and $\s8$. There are two additional cosmological priors, $h_0$
and $n_s$, which enter into the calculation of the linear matter power
spectrum for a given cosmological model, for which we use the transfer
function of \cite{eisenstein_hu:99}. We enforce Gaussian priors on
these parameters that are somewhat broader than their current
uncertainties considering the latest CMB results
(\citealt{komatsu_etal:10}). As we will demonstrate, the influence of
these parameters is minimal, but we include them for the purpose of
marginalizing over acceptable shapes of the matter power
spectrum. There are 7 additional priors on our systematic errors; 3
for dark matter halo statistics, 3 for the cluster mass-richness
relation, and one---$\fsys$---to incorporate systematic uncertainties
in our measurements.

The $\chi^2$ for the best-fit model is 26.9. With 16 free parameters
for 38 data points, this model yields a $\chi^2$ per degree of freedom
of 1.22. The probability of drawing a value $\ge 26.9$ from a $\chi^2$
distribution with 22 degrees of freedom is 0.215. Figure
\ref{bestfit_wp} shows the projected correlation function measured
from DR7 and the best-fit model. Due to the vanishingly small error
bars, we also show the fractional differences between the model and
data in Figure \ref{wpdiff}. The shaded bands show the 68\% and 95\%
ranges in the models from the chain at each value of $r_p$. Figure
\ref{bestfit_m2n} presents the maxBCG $M/N$ measurements and the
results from the MCMC chains. In each panel, the curve shows the
best-fit model, while the shaded regions once again indicate the 68\%
and 95\% ranges at each value of $\N$. The breakdown of the total
$\chi^2$ by the different data sets is as follows: $\chi^2$=6.1, 6.5,
and 5.7 for the $\mrz<-19.5$, $-20.5$, and $-21.0$ clustering samples,
respectively. For the $M/N$ data the $\chi^2=10.6$. Parameter values
and uncertainties for the HOD parameters are given in Table 4, while
cosmological constraints under various priors and assumptions are
listed in Table 5. We note that the best-fit value of $\fsys$ is
1.13---given our model assumptions, fitting the data favors $M/N$
values 18\% higher than our direct estimates, but within our estimated
systematic uncertainty.

\subsection{Cosmological Parameter Constraints}

Figure \ref{banana1} shows the main cosmological constraints in the
$\om$-$\s8$ plane. In cluster abundance studies, there is a natural
``banana curve'' degeneracy between $\om$ and $\s8$; the number of
clusters above a fixed mass varies with the overall matter density,
while the number of massive objects is highly sensitive to the
amplitude of matter fluctuations. This degeneracy curve also exists
for cluster $M/L$ or $M/N$ ratios, for the reasons described in \S
\ref{s.hod_demo}. The shaded banana curve in Figure \ref{banana1}
shows the results of the \cite{tinker_etal:05} analysis. The
degeneracy between $\om$ and $\s8$ is still seen in our analysis,
with a combined constraint of $\om^{0.5}\s8=0.465\pm 0.026$. The power
on $\om$ was chosen such that the likelihood of the parameter
combination was symmetric and yielded the smallest fractional
error. Without adding any additional datasets and marginalizing over
all other parameters, the constraints on individual cosmological
parameters are: $\om=0.29\pm 0.03$, $\s8=0.85\pm 0.06$ (68\%).  We
emphasize that these constraints follow from SDSS galaxy clustering
and maxBCG galaxy and weak lensing profiles, given only loose priors
on the primordial power spectrum and the broad assumptions of our HOD
framework.  Combining these results with the latest constraints from
WMAP7 (CMB-only constraints, assuming flat \lcdm;
\citealt{komatsu_etal:10}), our constraints are $\om=0.280\pm 0.012$,
$\s8 = 0.812\pm 0.016$.

Although the \cite{tinker_etal:05} result overlaps with our $1\sigma$
error contour in the $\om$-$\s8$ plane, the best-fit value of the
cluster normalization is higher in our current results. This offset is
driven primarily by the differences in the amplitude of the $M/N$
measurements from the maxBCG catalog relative to the mean $M/L$ ratio
of the 17 clusters in the CNOC2 survey. However, there are several
marked differences in the present analysis. First, we model $M/N$ as a
function of cluster richness, as opposed to using the mean $M/L$
ratio. Second, in \cite{tinker_etal:05} we fixed $\asat=1$. In this
paper, we find $1.0\lesssim \asat \lesssim 1.1$ depending on
luminosity.  If we had fixed $\asat=1.1$ in the previous analysis,
this would yield a $~6\%$ increase in the $\om^{0.6}\s8$ constraint
(see Fig. 4 in \citealt{tinker_etal:05}).  Lastly, systematic errors
in the HOD modeling, as well as redshift evolution from the mean CNOC2
redshift $z\sim 0.3$, were not included in the \cite{tinker_etal:05}
analysis. This would increase the errors and bring the two analyses
into better statistical agreement.

Figure \ref{constraints} illustrates the origin of the constraints in
the $\om$-$\s8$ plan, using data from the $\mrz<-20.5$ sample. We
choose four models from the MCMC chain that lie on our $2\sigma$
contour, two that bracket the width of the $\om^{0.5}\s8$ degeneracy
axis (both $\s8$ and $\om$ are low or high) and two that bracket the
range of allowed models along a constant $\om^{0.5}\s8$. Figure
\ref{constraints}b plots that fractional discrepancies of each model
with the $\wp$ data, while Figure \ref{constraints}c compares the
predicted and observed $M/N$. The $M/N$ agreement appears acceptable
in all cases, but that is because the $M/N$ normalization systematic
$\fsys$ is allowed to vary. For the high and low $\om^{0.5}\s8$
models, $\fsys$ has been forced to unlikely values that account for
much of the $\chi^2$. Figure 9d shows the $M/N$ predictions when
$\fsys$ is fixed to our best-fit value of 1.13. In this case, the high
and low $\om^{0.5}\s8$ models predict $M/N$ values that are clearly
too high and too low, respectively. The two models that define the
allowed range {\it along} the degeneracy axes are ruled out largely
by their discrepancies with the clustering data, especially for the
low-$\om$, high-$\s8$ model. Somewhat counter-intuitively, the
discrepancies with $\wp$ have the same sign in both cases, but the
$M/N$ data are also constraining the HOD---in the low-$\s8$ model,
lowering $\nsat$ at high masses would reduce $\wp$, but it would also
raise $M/N$ values are already too high.

Figure \ref{systematics_banana} demonstrates the effect of systematic
uncertainties on our cosmological constraints. In the first panel, our
uncertainties on the halo mass function, halo bias relation, and scale
dependent bias, are reduced to 1\%, 1\% and 3\%, respectively, which
dramatically shrinks the allowed parameter range along the
$\om^{0.5}\s8$ degeneracy axis. The tightening of the constraints is
due, in order of importance, to the reduced uncertainty in the
large-scale bias, the scale-dependent bias, and the mass function. The
uncertainty in the mass function has nearly negligible effect on the
results, which demonstrates once again the distinction between $M/N$
and cluster abundance constraints. The middle panel shows the effect
of the uncertainty on the systematics in the measurements---the
evolution of the luminosity function, the HOD, and the weak lensing
calibration, encapsulated by $\fsys$. Here we reduce the combined
uncertainties to 1\%, which leaves the range along the degeneracy
curve unchanged but shrinks the width of the $\om$-$\s8$ degeneracy
curve by about a factor of two. We note that once the prior on $\fsys$
goes below 5\%, the results are unchanged. At that level of precision,
we are statistically limited by the $M/N$ signal measured from the
cluster sample. The far right panel shows the cosmological constraints
when reducing both the theoretical and measurement uncertainties. For
a given galaxy clustering measurement, the $M/N$ data set the
amplitude of the $\om$-$\s8$ degeneracy axis. The value of $\fsys$
varies the amplitude of $M/N$, thus widening the error contours
parallel to this axis. The shape and amplitude of the galaxy
clustering measurements constrain the length of the error contours
along the axis. By reducing the systematics, it would be possible to
tighten the constraints on $\s8$ and $\om$ to $\sim 3-5\%$ from this
technique alone, without any additional data sets (see the lower three
rows in Table 5).

Figure \ref{systematics_banana2} explores the effect of other priors
in our analysis. In the left panel, we enforce a 10\% Gaussian prior
on the mass-concentration relation for satellite galaxies. This has
minimal impact on the cosmological constraints, demonstrating the
insensitivity of this approach to the details of the spatial bias of
galaxies within clusters---our method is only sensitive to the mean
number of satellites. The middle panel shows the effect on $\om$-$\s8$
if we tighten our priors on $h_0$ and $n_s$, which has the effect of
narrowing the range of shapes of the linear matter power
spectrum. Once again, there is little effect on cosmology from the
shape of $P(k)$---our method is most sensitive to its amplitude. The
right panel assumes minimal error on the mass-richness relation and
its scatter. Here again the change in the cosmological constraints is
minimal.

Figure \ref{grid_sys} shows the constraints on cosmological parameters
as well as the marginalization parameters.  In these contour plots one
can see clearly the dependence of our cosmological parameters on
$\fsys$, $\eb$, and $\er$, as well as the lack of correlation with
$\en$, $h_0$, and $n_s$. Of our systematic uncertainties, $\eb$ (the
normalization of the halo bias relation) shows the strongest
degeneracies with the cosmological parameters $\om$ and $\s8$. The
uncertainty on halo bias anticorrelates with $\s8$. The clustering
data determines the parameter combination $b_{\rm gal}\s8$, thus the
same fit to $\wp$ can be achieved with a lower $\s8$ and an
artificially enhanced bias factor. Because the $M/N$ data determine
the $\om^{0.5}\s8$ combination, $\eb$ shows a positive correlation
with $\om$ to preserve the cluster normalization as $\s8$ varies. The
red curves plotted on top of the histograms represent the prior
applied to that parameter. The histograms for $h_0$, $n_s$, $\eb$, and
$\en$ are coincident, or nearly so, with their priors. The best-fit
marginalized values of $\er$ and $\fsys$ are $\sim -1.3\sigma$ and
$+0.6\sigma$ off their prior values, respectively. For $\er$, a value
below unity brings the scale-dependent bias formula closer to
scale-{\it in}dependent. At scales below $\sim 1$ \hmpc, this
increases the amplitude of the halo-halo clustering. The value of
$\er$ is positively correlated with $\s8$ through their effect on
$\wp$ at the transition scale between pairs of galaxies in distinct
halos and pairs of galaxies that exist within a single halo (commonly
referred to as the `one-two to two-halo transition region'). Pairs of
galaxies at the transition regime can come from both
sources---satellite galaxies in massive halos and galaxy-galaxy pairs
in smaller halos (that can be closer together). If two-halo clustering
is higher at $r_p\lesssim 1$ \hmpc, there is less need for massive
clusters to supply the pairs to fit $\wp$ at that scale. The offset
between the MCMC results for $\fsys$ and its prior indicate some
tension between the clustering measurements and the maxBCG $M/N$
data. Recall that $\fsys$ multiplies the model calculations of M/N; to
fit the data with $\fsys=1$ requires a higher $\sigma_8$ than is
preferred by the clustering measurements.  However this offset is
within our uncertainty of the evolution between the redshifts of the
two samples. Although circumstantial, it is worth noting that the best
fit value of $\fsys$ is consistent with the amount and direction of
evolution in the HOD seen in the \cite{bower_etal:06} semi-analytic
galaxy formation model (see Appendix B), and with the direction of the
HOD evolution seen in our abundance matching models.

%%%%%%%%%%%%%%%%%%%%%%%%%%%%%%%%%
% Table 4
%%%%%%%%%%%%%%%%%%%%%%%%%%%%%%%%%
\begin{deluxetable*}{cccc}
  \tablecolumns{4} 
  \tablewidth{25pc} 
  \tablecaption{HOD Parameters Constraints}
  \tablehead{\colhead{Name} & \colhead{$\mrz<-19.5$} & \colhead{$\mrz<-20.5$} &  \colhead{$\mrz<-21.5$} }

\startdata

$\log\mmin$ & $11.59\pm 0.07$& $12.21\pm 0.11$ & $12.87\pm 0.12$ \\
$\log\msat$ & $12.94\pm 0.06$ & $13.46\pm 0.05$ & $13.87\pm 0.05$ \\
$\asat$ & $1.01\pm{0.04}$ & $1.03\pm 0.05$ & $1.08\pm 0.05$ \\
$\log\mcut$ & $12.48^{+0.31}_{-0.24}$& $12.60^{+0.25}_{-0.29}$ & $10.29^{+0.38}_{-0.89}$ \\
$\log\slogm$ & $-0.69$ & $-0.54^{+0.40}_{-0.71}$ & $-0.12^{+0.06}_{-0.08}$ \\
$\fcon$ & $1.61^{+0.34}_{-0.44}$& $0.97^{+0.45}_{-0.23}$ & $0.81^{+1.14}_{-0.55}$ \\
$\fsat$ & $0.16\pm 0.02$ & $0.15\pm 0.01$ & $0.13\pm 0.01$ \\

\enddata

\tablecomments{ All halo masses are in units of \hmsol. All logarithms
  are in base-10. $\fsat$ is the fraction of galaxies that are
  satellites. We note that any comparison between these results and
  others should account for any difference in halo mass definition.}
\end{deluxetable*}
%%%%%%%%%%%%%%%%%%%%%%%%%%%%%%%%%
% Table 4
%%%%%%%%%%%%%%%%%%%%%%%%%%%%%%%%%

%%%%%%%%%%%%%%%%%%%%%%%%%%%%%%%%%
% Table 5
%%%%%%%%%%%%%%%%%%%%%%%%%%%%%%%%%
\begin{deluxetable*}{lccc}
  \tablecolumns{3} 
  \tablewidth{25pc} 
  \tablecaption{Cosmological Parameter Constraints}
  \tablehead{\colhead{Name} & \colhead{$\om$} & \colhead{$\s8$} &\colhead{$\om^{0.5}\s8$} }

\startdata

$M/N$ & $0.29\pm 0.03$ & $0.85\pm 0.06$ & $0.465 \pm 0.026$\\
$M/N$+counts & $0.280^{+0.020}_{-0.023}$ & $0.809^{+0.044}_{-0.027}$&  \\
$M/N$+WMAP7 & $0.290\pm 0.016$ & $0.826\pm 0.020$ & \\
$M/N$+counts+WMAP7 & $0.280\pm 0.012$ & $0.812\pm 0.016$ &  \\
$M/N$+WMAP7+BAO+$H_0$ & $0.282\pm 0.011$ & $0.819\pm0.015$ & \\ 
%\hline\\
\hline\vspace{-0.1cm}\\
& & & $\om^{0.5}\s8$ \vspace{0.1cm}\\
$M/N~$ minus theory sys. & $0.276^{+0.026}_{-0.017}$ &  $0.908^{+0.045}_{-0.031}$ &  \\
$M/N~$ minus data sys. & $0.331^{+0.026}_{-0.036}$ & $0.842^{+0.050}_{-0.041}$ & $0.493^{+0.029}_{-0.010}$ \\
$M/N~$ minus all sys. & $0.297^{+0.019}_{-0.012}$ &  $0.952^{+0.029}_{-0.030}$ &\\

\enddata

\tablecomments{ Results including ``counts'' refer to the results from
  \cite{rozo_etal:10}. Results using WMAP7 data refer to seven-year
  CMB data from \cite{komatsu_etal:10}. These results assume a flat
  \lcdm\ cosmology. WMAP7 and WMAP7+BA0+$H_0$ results are taken from
  the publicly available MCMC chains from {\tt
    http://lambda.gsfc.nasa.gov/}. The bottom three parameter sets
  show constraints by reducing the errors on our systematics, as in
  Figure \ref{systematics_banana} (no additional data are
  included). `Minus theory' means setting the priors on $\en$, $\eb$, and
  $\er$ to 1\%, 1\%, and 3\%, respectively. `Minus data' means setting the
  error on $\fsys$ to 3\%. `Minus all' means combining both theory and data
  results.}
\end{deluxetable*}
%%%%%%%%%%%%%%%%%%%%%%%%%%%%%%%%%
% Table 5
%%%%%%%%%%%%%%%%%%%%%%%%%%%%%%%%%

\subsection{Halo Occupation Constraints}

The $1\sigma$ parameter constraints for all three clustering
samples are listed in Table 4. Figure \ref{grid_19.5} shows the
constraints on the HOD parameters for the $\mrz<-19.5$ clustering
sample. There is a strong degeneracy between $\asat$ and $\msat$. As
$\msat$ increases, the number of galaxies in high mass halos
decreases, but this can be compensated for by increasing
$\asat$. Tilting the $\nsat$ power law has the effect of reducing the
number of satellite galaxies in halos with $M<\msat$. It is satellite
galaxies in these low-mass halos that provide the pairs in the
correlation function at the smallest scales probed by our $\wp$
measurement, $r_p\lesssim 0.3$ \hmpc. Thus to fit $\wp$, when $\asat$
and $\msat$ increase, $\mcut$ decreases. In the best-fit model for
this sample, satellite galaxies are more centrally concentrated than
the dark matter by a factor of 1.6, but the constraints on this
parameter are weak and not strongly correlated with any other HOD
parameter.

For $\mrz<-20.5$, shown in Figure \ref{grid_20.5}, similar
degeneracies are seen between $\asat$, $\msat$, and $\mcut$. There is
a tail in the likelihood function for $\mcut$ to low masses; when
$\mcut\lesssim\mmin$, $\mcut$ no longer has an effect on the HOD or
$\wp$, thus all values below this mass scale are equally
likely. Rather than place a prior on $\mcut$ that depends on the value
of $\mmin$, we allow the chain to roam free through this part of
parameter space, which yields unusual likelihood shapes. For this
sample, values of $\slogm>1$ are strongly excluded because larger values of
the scatter decrease the large-scale bias of the sample and cannot be
reconciled with the observations even given the range of $\s8$ values
probed in the MCMC chain. For this sample, the spatial distribution of
the satellite galaxies is best represented by the concentration-mass
relation adopted for the dark matter halos, but the constraints on
$\fcon$ are also weak and do not depend on the other HOD parameters.

For the bright sample, shown in Figure \ref{grid_20.5}, $\slogm$ is
more tightly constrained; low values of the scatter produce too high a
large-scale bias relative to the $\wp$ measurements. A value of
$\slogm$ that monotonically increases with galaxy luminosity is
expected if the scatter in galaxy luminosity {\it at fixed halo mass}
is a constant, as widely assumed. The strange likelihood function for
$\mcut$ is due to the fact that the $\mcut<\mmin$ for most of the
chain, as discussed above.  The $\fcon$ parameter is unconstrained in
this model; the error bars in the one-halo term are somewhat larger
because of the smaller galaxy number density and the lower fraction of
galaxies that are satellites relative to fainter samples (see, e.g.,
Z10). 

For all three samples, the power-law index on the satellite occupation
function, $\asat$, is close to unity, as expected from theoretical
studies and from previous analyses of SDSS clustering. There is some
tension between our values of $\asat$ and those derived from the
clustering-only analysis of Z10. For $\mrz<-20.5$, and $\mrz<-21.0$,
Z10 find $\asat=1.15\pm 0.03$ and $\asat=1.15 \pm 0.06$,
respectively. Our analysis yields $\asat=1.03^{+0.04}_{-0.03}$ and
$\asat=1.08\pm 0.05$. Some of this difference may be due to the choice
of $\s8=0.8$ in Z10; this is $1\sigma$ below our best-fit value. But
to account for the full difference, the $M/N$ data must be
included. It is notable that our constraints on HOD parameters are not
appreciably better than those of Z10 even with the addition of extra
data on the HOD; however, Z10 fix their cosmology, which naturally
tightens the constraints on the HOD parameters, and the include larger
scale $\wp$ data. Z10 use a cutoff power-law to parameterize $\nsat$,
as opposed to the exponential cutoff in eq. (\ref{e.nsat}), and they
assume $\Delta_{\rm halo}=200\rho_m$ for all calculations, thus a
one-to-one comparison of the two is not straightforward.

\begin{figure*}
\epsscale{1.0} 
%\plotone{banana_zoom.ps}
\plottwo{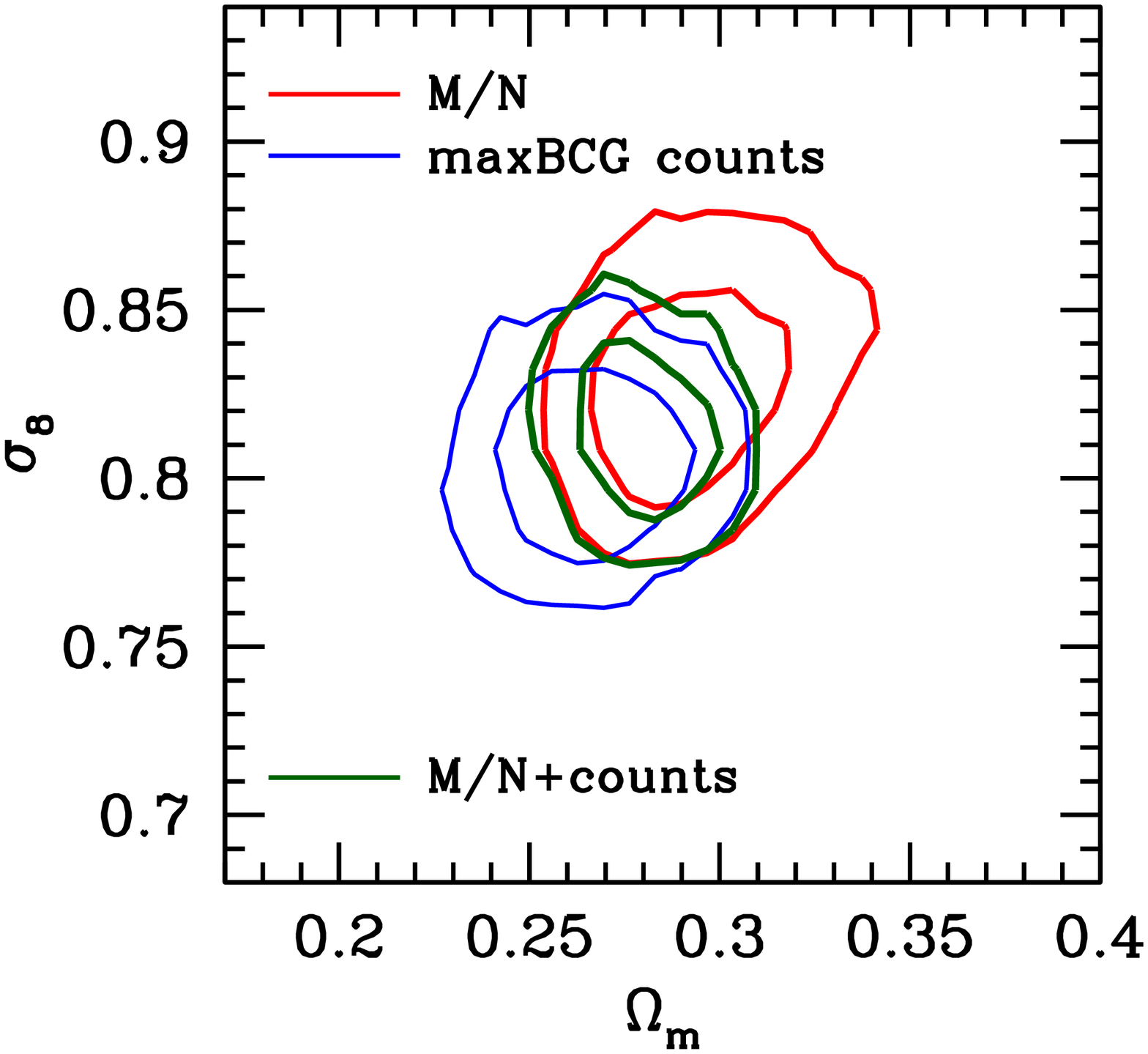}{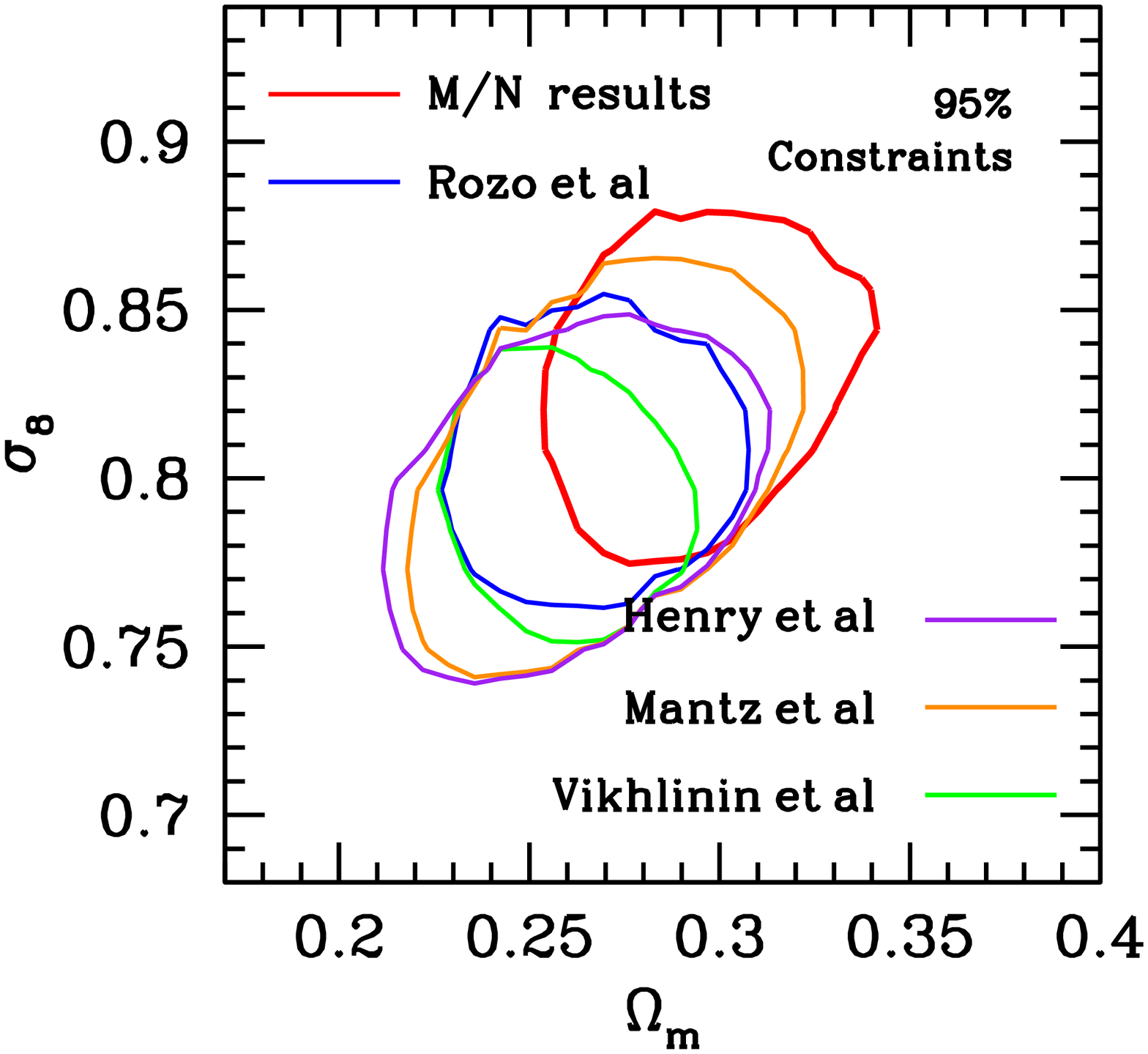}
%\vspace{-7.5cm}
\caption{ \label{banana_zoom} {\it Left Panel}: Comparison of the
  $M/N$ results to parameter constraints from maxBCG cluster counts
  from \cite{rozo_etal:10}. Both results assume WMAP7 priors. It is
  important to note that cluster abundances are not used anywhere in
  our analysis. The systematic uncertainties that Rozo et~al.~are most
  sensitive to are the ones on which $M/N$ have little
  dependence. Thus, the constraints from these two studies are roughly
  independent. {\it Right Panel}:Comparison of this work to other
  cluster abundance studies. All contours are 95\% and incorporate
  WMAP7 priors. The $M/N$ technique is consistent and competitive with
  cluster abundance studies while using complementary information to
  cluster counts.}
\end{figure*}

\section{Discussion}

\subsection{Comparison to Cluster Abundance Studies}

Throughout the analysis in this paper, the information from cluster
number counts---i.e., the space density of clusters as a function of
richness or mass---has not been used. Figure \ref{banana_zoom}
compares our results to those of various cluster abundance studies. In
particular, the left-hand side compares the cosmological constraints
from $M/N$ to those of abundances from maxBCG clusters
(\citealt{rozo_etal:10}). Both of these analyses use the same sample
of clusters, but in nearly complementary ways. The $M/N$ results are
both in good agreement with the abundance constraints and
competitive with the abundance constraints. The right-hand side of
Figure \ref{banana_zoom} presents the 95\% confidence region in the
$\om$-$\s8$ plane for the two maxBCG studies in comparison to several
cluster abundance results from X-ray observations. Here again, the
$M/N$ results are in good agreement with and competitive with these
approaches.

There are several ways in which the $M/N$ approach is complementary to
and independent of the cluster abundance method, even when using the
same cluster sample. The parameters $\er$, $\eb$, and the evolution of
the luminosity function have no impact on abundance analysis. The
dominant theoretical systematic uncertainty for cluster abundances is
the shape and normalization of the halo mass function. From Figure
\ref{grid_sys}, the dependence of cosmology on this uncertainty is
nearly negligible. In the analysis of maxBCG counts by
\cite{rozo_etal:10}, the dominant systematic uncertainties are
miscentering and the normalization of the weak lensing mass
calibration. Miscentering has less of an effect on $M/N$ than on
abundance analysis. The weak lensing measurements of
\cite{sheldon_etal:09_data} and halo mass estimates of
\cite{johnston_etal:07} are used to calibrate the mass-richness
relation in the \cite{rozo_etal:10} results, thus the uncertainty in
the weak-lensing calibration is correlated between the $M/N$ and
abundance analyses. However, this uncertainty is sub-dominant in the
$M/N$ analysis. Removal of this uncertainty entirely would only change
the prior on $\fsys$ from 0.227 to 0.225. A larger error in the weak
lensing calibration than that suggested by \cite{rozo_etal:10} would
affect both measurements. A fully robust analysis of the maxBCG
cluster catalog should simultaneously model $N(\N)$, $\wp$, and $M/N$,
but given the different dependences of the techniques on the common
systematic uncertainties, the two results can be considered nearly
independent. Combined constraints using $M/N$, cluster counts, and CMB
data are listed in Table 5. For non-optical cluster counts, such as
detections through X-ray or Sunyaev-Zeldovich observations, the
dominant observational systematics are the absolute normalization and
scatter of the mass-observable relation. Because $\asat$ is close to
unity for all galaxy samples, the $M/N$ function is nearly independent
of cluster richness and thus of scatter in mass versus
richness. Figure \ref{systematics_banana2} demonstrates the
insensitivity of our results to our assumptions about the uncertainty
in this scatter.

If the theoretical uncertainties on $\en$, $\eb$, and $\eb$ were
significantly smaller, our best fit value for $\s8$ would be $\sim
0.91$. This would be in tension with the WMAP7 cosmological
constraints as well as cluster abundance results. This tension could
be relieved if the true halo bias function had a higher normalization
than the \cite{tinker_etal:10_bias} result by $\sim 3\%$, or a
scale-dependent bias that is closer to scale-independent relative to
the non-linear dark matter clustering.

\subsection{Prospects for the Future}

Substantial improvements in the cosmological constraints presented in
this paper can be achieved by reducing the systematic uncertainties in
the current analysis. The theoretical uncertainties most applicable to
our analysis, namely the large-scale bias and scale-dependent bias of
dark matter halos, can be reduced with improved numerical
simulations. Although state-of-the-art just two years ago, the N-body
simulation set analyzed in \cite{tinker_etal:08_mf} and
\cite{tinker_etal:10_bias} has been replicated many times over (in
volume and particle number) by ongoing numerical studies. The
simulations of the Las Damas project\footnote{\tt
  http://lss.phy.vanderbilt.edu/lasdamas/} can currently achieve
marked improvement in the halo mass function (McBride et~al., in
preparation) and the bias statistics for typical \lcdm\ cosmologies. A
fully robust theoretical framework requires understanding of halo
populations in dark energy cosmologies, although we expect the impact
of dark energy to be small at fixed $\om$, $\s8$, and power spectrum
shape (e.g., \citealt{kuhlen_etal:05}).

For the systematic uncertainties in our measurements, there exists a
simple solution to remove the uncertainty on the evolution of the
luminosity function and the HOD between the clustering samples and the
sample of clusters: measure the clustering from the same volume of
space as the maxBCG sample. A spectroscopic sample of luminous red
galaxies (LRGs) with $\ngavg \approx 1\times 10^{-4}$ extends to
$z=0.36$ (see, e.g., \citealt{kazin_etal:10}). Nearly every maxBCG
cluster above a richness limit of $\N=9$ ($M\approx 3\times 10^{13}$
\hmsol) should contain one or more LRGs (\citealt{zheng_etal:10}),
thus this analysis can be repeated on a sample of LRGs with an
uncertainty on $\fsys$ than only comes from the weak lensing
calibration. From the low number density of LRGs, as well as the halo
occupation analysis of \cite{zheng_etal:10}, we expect $\nsat$ for
LRGs to be roughly $1/4$ that of the bright sample analyzed here,
increasing the statistical errors on $\nsat$ from $\sim 5\%$ to
$10\%$, but the size of the maxBCG cluster sample analyzed can be
increased by a factor of two relative to the sample used in
\cite{sheldon_etal:09_data} and in this paper. In the near future, the
Baryon Oscillation Spectroscopic Survey (BOSS; \citealt{boss}) will
increase the number density of LRGs at $z\lesssim 0.4$ by a factor of
three relative to the DR7 LRG number density. All of this points to
luminous red galaxies as a logical and straightforward extension of
the $M/N$ analysis.

In the longer term, there are numerous photometric surveys planned
that will produce vast numbers of optically-detected galaxy clusters
out to $z\gtrsim 1$, including the Dark Energy Survey, Pan-STARRS, and
the Large Synoptic Survey Telescope. These surveys are also designed
for weak lensing in order to obtain cluster masses as well as cosmic
shear to measure the amplitude of dark matter clustering.  These
surveys will yield precise measurements of the angular clustering of
$\lstar$ galaxies. The $M/N$ technique can be applied to these data
with little modification from the analysis presented here. Due to the
complementary nature of the $M/N$ approach to cluster abundances, this
technique will enhance the measurements on the growth and expansion
history of the universe and tighten constraints on the equation of
state of dark energy.  For photometric data, the dominant systematic
will be uncertainties in the photometric redshift errors, and it will
be important to quantify how the effect of photometric redshift
uncertainty correlates $M/N$ and cluster abundance data. As opposed to
cluster counts, which are sensitive both to the growth of structure
and to the expansion history of the universe, the $M/N$ approach is
primarily sensitive to growth because the volume element does not
enter directly into the $M/N$ predictions. Thus the combination of
these two approaches may break the degeneracy between dark energy and
models in which gravity is modified at large scales in order to
explain late-time acceleration (e.g., \citealt{knox_etal:06,
  huterer_linder:07}).

\section{Summary}

We have presented measurements of the mass-to-galaxy-number ratio
($M/N$) within maxBCG clusters in the SDSS. Combined with measurements
of the projected clustering of galaxies, we have demonstrated that
these data are a sensitive probe of the cosmological parameters $\om$
and $\s8$, while being insensitive to other cosmological parameters
that mainly enter into the shape of the linear matter power
spectrum. Our theoretical analysis is based on the halo occupation
distribution (HOD). We have incorporated uncertainties in the
theoretical modeling, including uncertainties in the halo mass
function, the halo bias function, and the scale-dependent bias of
halos. We are most sensitive to uncertainties in the halo bias
function and to the possibility of evolution between the mean redshift
of the cluster sample, $z\sim 0.25$, and the mean redshift of the
clustering sample, $z\sim 0.1$. After incorporating all these
uncertainties in our analysis, we find $\om^{0.5}\s8=0.465\pm 0.026$,
with individual constraints of $\om=0.29\pm 0.03$ and $\s8=0.85\pm
0.06$. Combined with current CMB data, these constraints are
$\om=0.290\pm 0.016$ and $\s8=0.826\pm 0.020$. These constraints are
consistent with and comparable to those obtained by cluster
abundances, even though abundance information is not used in this
analysis. Future investigations could reduce the systematic
uncertainties in this analysis and thereby tighten parameter
constraints by a factor of two or more.

The systematic uncertainties that are most important for cluster
abundance studies, namely theoretical uncertainties in the halo mass
function and scatter in the mass-observable relation, have negligible
impact on the constraints achieved with $M/N$. Thus the combination of
these two techniques can provide a unique probe of the growth and
expansion history of the universe from future photometric galaxy
surveys.

\acknowledgements

RHW received support from the DOE under contract DE-
AC03-76SF00515. MTB and RHW thank their collaborators on the LasDamas
project for critical input on the Carmen simulation, which was
performed on the Orange cluster at SLAC. ER is funded by NASA through
the Einstein Fellowship Pro- gram, grant PF9-00068. DHW acknowledges
the supprt of NSF grant AST-1009505. IZ acknowledges support by NSF
grant AST-0907947.

\appendix
\section{A. Modified Scale-Dependent Bias}

In this paper we use a modified form of the scale-dependent bias
function presented in \cite{tinker_etal:05}. That function was
calibrated on a series of N-body simulations in which the halos were
identified using the friends-of-friends percolation algorithm (FOF;
e.g., \citealt{davis_etal:85}), with a linking length of 0.2 times the
mean interparticle separation. The halo mass function and halo bias
function used here use the spherical overdensity algorithm (SO;
\citealt{tinker_etal:08_mf, tinker_etal:10_bias}). In FOF, nearby
halos can be linked together and labeled as a single object. In SO,
this rarely happens. In the SO algorithm implemented by Tinker
et.~al., halos are allowed to overlap so long as the center of one
halo is not within the virial radius of another. Thus the small-scale
clustering of halos in these two scenarios should not be the same. For
pair separations $r\ge R_1+R_2$, the \cite{tinker_etal:05} function is
an adequate description of the scale dependence of bias. For pairs in
the overlap regime, $r<R_1+R_2$, we find that the bias with respect to
the non-linear matter distribution is nearly constant. Thus we adopt

\begin{equation}
\label{e.scale_bias}
  b^2(M,r)= \left\{ \begin{array}{ll}
      b^2(M)\, \frac{[1+1.17\,\xi_m(r)]^{1.49}}{[1+0.69\,\xi_m(r)]^{2.09}} & {\rm if\ \ } r>=2R_{\rm halo} \\
      b^2(M)\, \frac{[1+1.17\,\xi_m(2R_{\rm halo})]^{1.49}}{[1+0.69\,\xi_m(2R_{\rm halo})]^{2.09}} & {\rm if\ \ } r<2R_{\rm halo} \\
        \end{array}
        \right.
\end{equation}

\noindent as our functional form. At $r<R_{\rm halo}$, $b(M,r)=0$ by
definition, but this is enforced in the halo exclusion of the two-halo
term. See Appendix B of \cite{tinker_etal:05} for full details on the
analytic model, with the minor differences described in \S
\ref{s.hod_model}.

\begin{figure}
\epsscale{0.8} 
\plotone{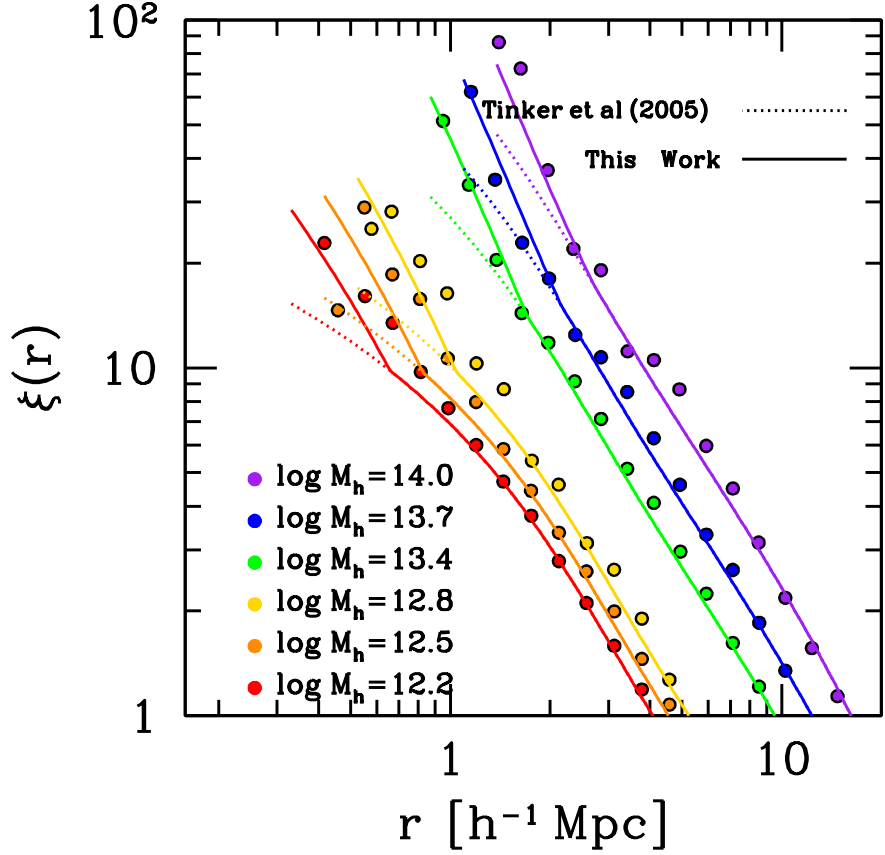}
% from REDSHIFT_DISTORTIONS/HALO_BIAS/sm.scalebias
%\vspace{-7.5cm}
\caption{ \label{scalebias} Halo-halo correlation functions from two
  high resolution N-body simulations (the L1000W and H384 simulations
  utilized to calibrate both the halo mass function and halo bias
  relation in \citealt{tinker_etal:08_mf} and
  \citealt{tinker_etal:10_bias}). The three low-mass bins are from the
  H384 simulation and the three high-mass bins are from the L1000W
  simulation. The abrupt increase in clustering amplitude between the
  two simulation results is due to the change in cosmological
  model. The dotted line shows the scale-dependent bias function of
  \cite{tinker_etal:05}. The solid curves show our modification to
  this formula to account for the change in halo definition between
  the \cite{tinker_etal:05} simulations and the more recent
  calibrations. See text for details.}
\end{figure}

Figure \ref{scalebias} shows the halo autocorrelation functions for
five mass bins. The data points are taken from two simulations
described in \cite{tinker_etal:10_bias}: L1000W and H384. Results are
shown here for halos with $N\ge 400$ particles per halo. The curves
are calculated assuming $\xi(r)=b^2(M,r)\xi_m(r)$, where $\xi_m(r)$ is
the non-linear matter correlation function given by
\cite{smith_etal:03}. The dotted curves show the original function and
the solid curves show the modification in eq. (\ref{e.scale_bias}). It
is clear that eq. (\ref{e.scale_bias}) is a better description of
halo-halo clustering at small scales. The fit is not perfect, however,
which is why we have adopted a 15\% error on the deviation of the bias
from scale independence.

\section{B. Evolution of the HOD in Semi-Analytic Models}

\begin{figure}
\vspace{-11cm}
\epsscale{1.25} 
\plotone{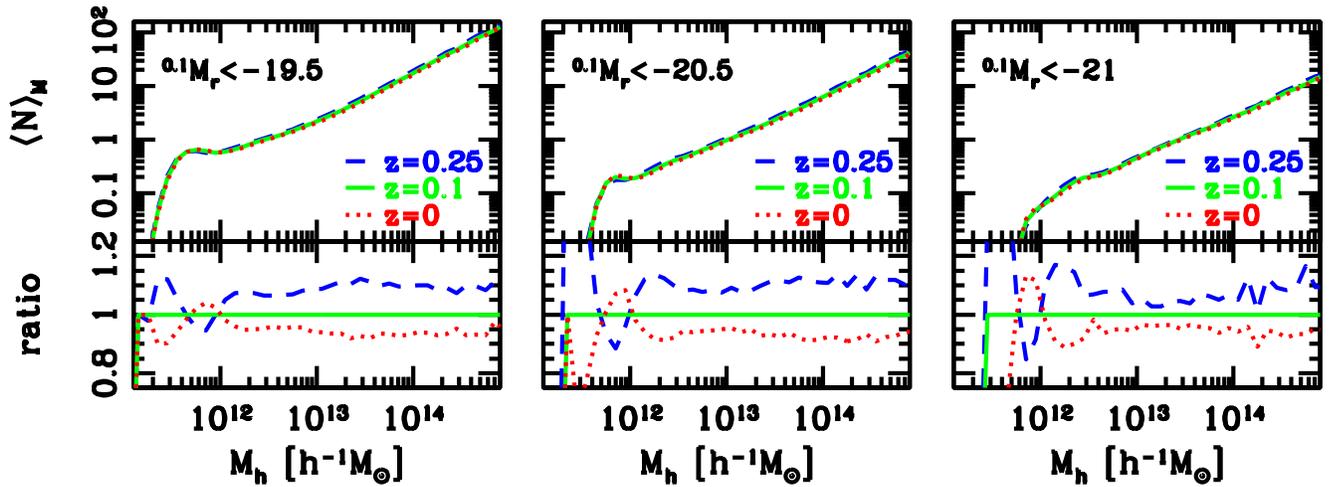}
\caption{ \label{hod_evol} Evolution of the HOD in the
  \cite{bower_etal:06} semi-analytic galaxy formation model. The right,
  middle, and left panels show results for galaxy samples that match
  the number densities of our $\mrz<-21$, $\mrz<-20.5$, and
  $\mrz<-19.5$ samples, respectively. The lower frame in each panel
  shows the ratio with respect to the $z=0.1$ HOD.  }
\end{figure}

One of our primary uncertainties is whether the HOD measured at
$z=0.25$ should be the same as that inferred from clustering at
$z=0.1$. In a separate paper we will demonstrate that the halo
occupation of galaxies inferred through the subhalo abundance matching
paradigm shows negligible evolution for samples defined at a fixed
number density (Reddick et al., in preparation). We have also
investigated the evolution of the HOD in the semi-analytic galaxy
formation model of \cite{bower_etal:06}, built upon the high
resolution Millennium N-body simulation
(\citealt{springel_etal:05}). This model compares well with the observed luminosity
and color distribution of galaxies, and it does a reasonable job
reproducing the dependence of galaxy properties on environment
(\citealt{baldry_etal:06}).  Figure \ref{hod_evol} shows the evolution
of the HOD for galaxies in the SA model for the same three number
density thresholds as our $z=0.1$ clustering samples. In each panel,
the upper frame shows the HOD at $z=0.25$, $z=0.1$ and $z=0$. The
lower frame shows the ratio of $\navg{_{z=0.25}}$ to $\navg$ at
$z=0.1$ and $z=0$. For the halo mass range probed by the maxBCG
sample, and over the redshift baseline of our observations, $\navg$
varies by 5-10\%, depending on the galaxy number density. We
incorporate a 10\% uncertainty in the evolution of the HOD in our
analysis.

\section{C. Tests with Mock Galaxies}

We have tested our methodology on mock galaxy distributions created on
a high-resolution N-body simulation. The simulation is H384 from the
simulation set analyzed in \cite{tinker_etal:08_mf,
  tinker_etal:10_bias}. The simulation consists of 1024$^3$ particles
evolved in a volume 384 \hmpc\ per side. The mass resolution of this
simulation is high enough to resolve halos well below $\mmin$ of the
$\mrz<-19.5$ galaxy sample. The simulation cosmology is
$(\om,\s8,\omb,h_0,n_s)=(0.3,0.9 0.04, 0.7, 1.0)$. The halo catalogs
are the same as those in the Tinker et.~al.\ mass function and bias
relation analyses, using the SO halo finder.

Our procedure for creating the mock data was as follows. First, we fit
the $\wp$ data alone assuming the same cosmology as the
simulation. The halos within the simulation are populated with
galaxies according to the best-fit HODs. For each halo, a maxBCG
richness $\N$ is assigned according to the mass-richness relation
given in eq.~(\ref{e.mass_richness}) (including scatter). The halos
are stacked in the same richness bins given in Table 2, and the mean
value of $M/N$ for each bin is calculated directly from the set of
halos in each bin; i.e., we do not create shear profiles or projected
number density profiles from the galaxy distribution. Our test was
directed at the methodology of combining $\wp$ and $M/N$ itself,
rather than the techniques for measuring these quantities, which are
tested elsewhere and are included in the systematic error budget in
our analysis. We include all the same systematic uncertainties in this
analysis---halo statistics, mass-richness relation, cosmological
parameters---except for $\fsys$, which we fixed at unity and do not
allow to vary. We obtain the error bars on $\wp$ from performing the
jackknife technique on the simulation itself, but for $M/N$ we adopt
the same (proportional) errors as on the maxBCG data.

\begin{figure}
\epsscale{0.8} 
\plotone{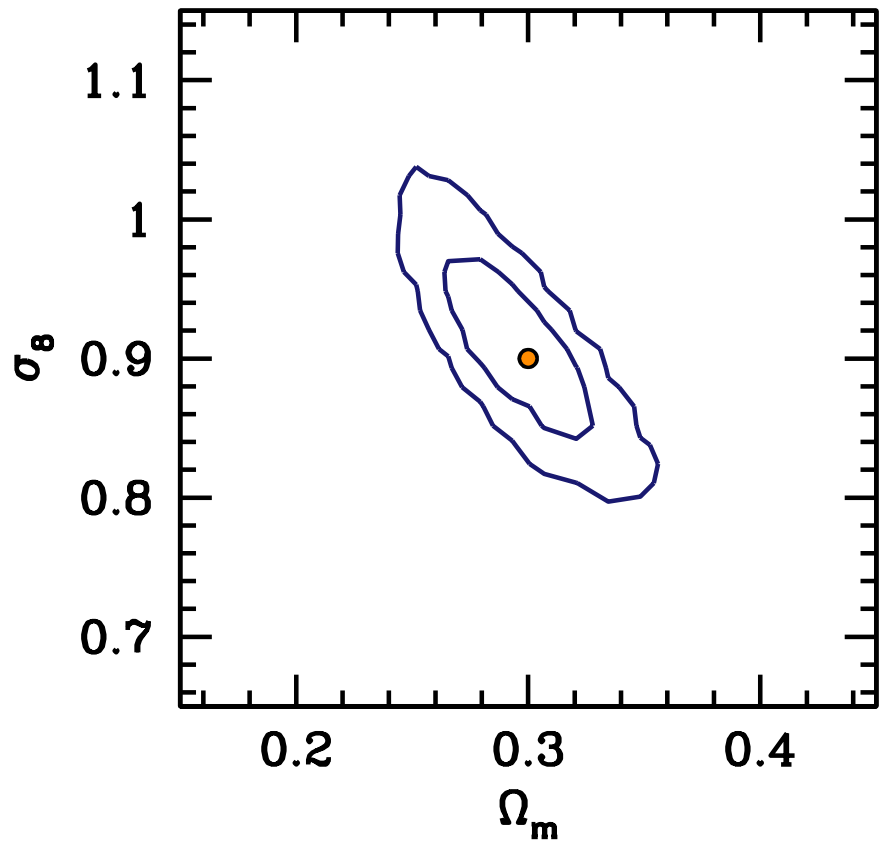}
%\vspace{-7.5cm}
\caption{ \label{mock_banana} Cosmological constraints on mock data
  created from an N-body simulation 384 \hmpc\ per side (the H384
  simulation utilized to calibrate both the halo mass function and
  halo bias relation in \citealt{tinker_etal:08_mf} and
  \citealt{tinker_etal:10_bias}). The mock analysis incorporates
  uncertainties on $\en$, $\er$, and $\eb$, but assume that $\fsys=1$
  with no error. The circle indicates the input cosmology.  }
\end{figure}

Figure \ref{mock_banana} shows the cosmological constraints produced
by the mock analysis. The error contours are centered on the input
cosmology. The banana curve is similar to the results from the actual
data but with a stronger constraint on the cluster normalization due
to the tight prior on $\fsys$.

%%%%%%%%%%%%%%%%%%%%%%%%%%%%%%%%%%%%%%%
% These are figures for the appendices
%%%%%%%%%%%%%%%%%%%%%%%%%%%%%%%%%%%%%%%

%%%%%%%%%%%%%%%%%%%%%%%%%%%%%%%%%%%%%%%%%%%%%%%%%%%%%%%%%%%%%%%%%%%%%%%%
%  Bibliography
%%%%%%%%%%%%%%%%%%%%%%%%%%%%%%%%%%%%%%%%%%%%%%%%%%%%%%%%%%%%%%%%%%%%%%%%

\bibliography{../risa}

\end{document}